\documentclass[sigconf]{acmart}

\usepackage{pifont}
\usepackage{amsfonts}
\usepackage{zlmtt}			%
\usepackage{amsmath}
\usepackage[nointegrals]{wasysym}
\usepackage{mathtools}
\usepackage{fontawesome}        %

\usepackage{soul}

\usepackage{alltt}

\usepackage{epstopdf}		%
\PassOptionsToPackage{usenames,dvipsnames,svgnames,table}{xcolor}
\definecolor{applegreen}{rgb}{0.55, 0.71, 0.0}
\definecolor{arylideyellow}{rgb}{0.91, 0.84, 0.42}
\definecolor{moonstoneblue}{rgb}{0.45, 0.66, 0.76}
\definecolor{azure}{rgb}{0.0, 0.5, 1.0}
\definecolor{blush}{rgb}{0.87, 0.36, 0.51}
\definecolor{bittersweet}{rgb}{1.0, 0.44, 0.37}
\definecolor{ForestGreen}{rgb}{0.0, 0.27, 0.13}
\definecolor{Thistle}{rgb}{0.85, 0.75, 0.85}

\colorlet{mgray}{gray!90!black}
\definecolor{codegreen}{rgb}{0,0.6,0}
\definecolor{codegray}{rgb}{0.5,0.5,0.5}
\definecolor{codepurple}{rgb}{0.58,0,0.82}
\definecolor{backcolour}{rgb}{0.95,0.95,0.92}

\usepackage{hyphenat}
\usepackage{hyperref}
\usepackage{xurl} %
\usepackage{hyperref}
\hypersetup{
    colorlinks = true,
    citecolor = {blue},
    linkcolor = {cyan},
}

\usepackage{xspace}		%
\usepackage{balance}	%

\usepackage{tikz}
\tikzset{mynode/.style={draw=blue,circle,inner sep=1pt,font=\scriptsize,anchor=south}
}
\usepackage{titlesec}

\usepackage[olditem,oldenum]{paralist} %
\usepackage[shortlabels]{enumitem} %

 \usepackage{array}
\usepackage{booktabs}		%
\usepackage{colortbl}		%
\usepackage{multicol}		%
\usepackage{multirow}		%
\usepackage{threeparttable} %

\usepackage{comment}

\usepackage{listings}

\lstdefinestyle{x86}{
  language={[x86masm]Assembler},   %
  title=\lstname,                 %
}

\lstdefinestyle{CStyle}{
  language=C,                %
  title=\lstname,                 %
}

\lstdefinestyle{C++Style}{
  language=C++,                %
  title=\lstname,                 %
}

\lstdefinestyle{bash}{
  language=bash,                %
  stepnumber=1,                   %
  numbersep=5pt,                  %
  backgroundcolor=\color{white},  %
  showspaces=false,               %
  showstringspaces=false,         %
  showtabs=false,                 %
  tabsize=2,                      %
  basicstyle=\footnotesize,
  breaklines=true,                %
  breakatwhitespace=true,         %
  title=\lstname,                 %
  keywordstyle=\color{blue},
  stringstyle=\color{purple},
  commentstyle=\color{gray}
}

\usepackage{algorithm}
\usepackage[noend]{algpseudocode}

\usepackage{verbatimbox}
\usepackage[graphicx]{realboxes}

\usepackage{lipsum} %

\usepackage{cleveref}
\crefformat{footnote}{#2\footnotemark[#1]#3}

\newcommand{\ie}{i.e.,\xspace}
\newcommand{\eg}{e.g.,\xspace}

\newcommand{\etc}{etc.\xspace}
\newcommand{\vs}{vs.\xspace}

\newcommand{\code}[1]{\texttt{#1}}

\def\cmark{\ding{51}} 
\def\xmark{\ding{55}} 

\def\asti{$^\ast$}
\def\astii{$^\dagger$}
\def\astiii{$^\diamond$}
\def\astiv{$^\ddagger$}

\newif\ifcomments
\commentstrue

\ifcomments
	\newcommand{\gp}[1]{{[\textcolor{purple}{GP: #1}]}}
	\newcommand{\vr}[1]{{[\textcolor{blue}{VR: #1}]}}
	\newcommand{\jx}[1]{{[\textcolor{green}{JX: #1}]}}
	\newcommand{\kk}[1]{{[\textcolor{orange}{KK: #1}]}}
\else
    \newcommand{\gp}[1]{}
    \newcommand{\vr}[1]{}
    \newcommand{\jx}[1]{}
    \newcommand{\kk}[1]{}
\fi

\newif\ifshowdiff

\ifshowdiff
\def\revcolor{blue}
\else
\def\revcolor{black}
\fi

\def\system{\textsc{SysPart}\xspace}
\def\sf{\textsc{sysfilter}\xspace}
\def\typearmor{\textsc{TypeArmor}\xspace}

\def\tsp{TSP\xspace}
\def\SF{SF\xspace}

\def\epatch{\textsc{e9patch}\xspace}

\def\nginx{{Nginx}\xspace}
\def\bind{{Bind}\xspace}
\def\httpd{{Httpd}\xspace}
\def\lighttpd{{Lighttpd}\xspace}
\def\memcached{{Memcached}\xspace}
\def\redis{{Redis}\xspace}

\def\mainloop{main loop\xspace}
\def\tspfixed{TSP\textsubscript{fixed}\xspace}
\def\seccomp{Seccomp-BPF\xspace}

\newcommand{\loopfunc}[1]{\code{{\small #1}}}
\newcommand{\badloopfunc}[1]{\code{\textcolor{red}{{\small #1}}}}

\setitemize{itemsep=0em,topsep=0em,parsep=0pt,partopsep=0pt,leftmargin=*}
\setenumerate{itemsep=0em,topsep=0em,parsep=0pt,partopsep=0pt,leftmargin=*}
\newlist{points}{itemize}{1}
\setlist[points,1]{label=$\blacktriangleright$,itemsep=0em,topsep=0em,parsep=0pt,partopsep=0pt,leftmargin=0pt, itemindent=*}

\titlespacing{\paragraph}{0pt}{0.3em}{0.5em}
\titleformat{\paragraph}[runin]{\normalfont\bfseries}{\theparagraph}{}{}

\copyrightyear{2023}
\acmYear{2023}
\setcopyright{acmlicensed}
\acmConference[CCS '23]{Proceedings of the 2023 ACM SIGSAC Conference on Computer and Communications Security}{November 26--30, 2023}{Copenhagen, Denmark}
\acmBooktitle{Proceedings of the 2023 ACM SIGSAC Conference on Computer and Communications Security (CCS '23), November 26--30, 2023, Copenhagen, Denmark}
\acmPrice{15.00}
\acmDOI{10.1145/3576915.3623207}
\acmISBN{979-8-4007-0050-7/23/11}
\begin{document}
\renewcommand{\abstractname}{ABSTRACT}
\title{\system: Automated Temporal System Call Filtering for Binaries}

\author{Vidya Lakshmi Rajagopalan}
\affiliation{%
  \institution{Stevens Institute of Technology}
  \city{Hoboken, NJ}
  \country{USA}}

\author{Konstantinos Kleftogiorgos}
\affiliation{%
  \institution{Stevens Institute of Technology}
  \city{Hoboken, NJ}
  \country{USA}}

\author{Enes G\"{o}kta\c{s}}
\affiliation{%
  \institution{Stevens Institute of Technology}
  \city{Hoboken, NJ}
  \country{USA}}

\author{Jun Xu}
\affiliation{%
  \institution{University of Utah}
  \city{Salt Lake City,UT}
  \country{USA}}

\author{Georgios Portokalidis}
  \affiliation{%
    \institution{Stevens Institute of Technology}
    \city{Hoboken, NJ}
    \country{USA}}
\affiliation{%
  \institution{IMDEA Software Institute}
  \city{Madrid}
  \country{Spain}}

\begin{abstract}

Restricting the system calls available to applications reduces the attack
surface of the kernel and limits the functionality available to compromised
applications. Recent approaches automatically identify the system calls
required by programs to block unneeded ones. For servers, they
even consider different phases of execution to tighten restrictions after
initialization completes. However, they require access to the source code for
applications and libraries, depend on users identifying when the server
transitions from initialization to serving clients, or do not account for
dynamically-loaded libraries.
This paper introduces \system, an automatic system-call filtering system
designed for binary-only server programs that addresses the above limitations.
Using a novel algorithm that combines static and
dynamic analysis, \system identifies the serving phases of all working threads of a server. Static analysis is
used to compute the system calls required during the various serving phases in
a sound manner, and dynamic observations are only used to complement static
resolution of dynamically-loaded libraries when necessary.
We evaluated \system using six popular servers on x86-64 Linux to demonstrate
its effectiveness in automatically identifying serving phases, generating
accurate system-call filters, and mitigating attacks. Our results show that
\system outperforms prior binary-only approaches and performs comparably to
source-code approaches.

\end{abstract}



\maketitle

\section{INTRODUCTION}
\label{sec:intro}

Applications interact with operating systems (OSs) through system calls
(syscalls). Over time, the number of syscalls has increased to accommodate the
growing size and complexity of application software. Indicatively, the latest
Linux kernel (v6.3.1) provides 451 syscalls in x86-64 architectures, compared
to just 347 in the previous version (v5.5). Astoundingly, more than 100 new
syscalls have been added since then. However, this increase is not without
risks. Unprivileged applications, whether malicious or
compromised~\cite{bluekeep}, can exploit vulnerabilities in system-call code,
which runs with higher privileges, to elevate their own
privileges~\cite{kemerlis2012kguard, kemerlis2014ret2dir, pomonis2017kr,
	pomonis2019kernel, kemerlis2015protecting, li2017lock}. System calls are also
the only means by which malicious applications can execute dangerous actions,
such as downloading and running malware, communicating over the network, and
so on.

\paragraph{The Literature:}

To mitigate this issue, recent approaches attempt to limit the number of
system calls available to applications. One such approach is
\sf~\cite{sysfilter} (\SF), which utilizes binary analysis to statically approximate
the minimum set of system calls required by an entire binary application over
its lifetime. A filter is then installed to block unnecessary calls at load
time. In contrast, temporal system call specialization~\cite{temporal} (\tsp)
focuses on server applications, and partitions the application lifetime into
initialization and serving phases. Static analysis at the compiler-level is
employed to further restrict the system calls available during the serving
phase. While \tsp offers superior security, it has several limitations: \ding{182} \tsp works only on source code and cannot be
applied to applications with binary components; \ding{183} it depends on the
manual identification of the serving phase; and \ding{184} it ignores
dynamically-loaded libraries (DLL), potentially resulting in false positives
when syscalls are needed for those libraries.

\paragraph{Our Approach:}

In this paper, we present \system, a binary-only,
automatic system-call filtering system for server programs. We introduce two
new techniques to overcome \tsp's limitations.

\begin{points}
\item
Similar to \tsp, we divide a server's lifetime into two phases: initialization
and serving. However, unlike \tsp, \system automatically identifies the
beginning of the serving phase. Our approach employs a novel algorithm that
combines static and dynamic analysis with simple workloads, freeing the user
from the burden of manual identification. We observe that a server's serving
phase typically employs a loop structure, and we identify a loop that
\emph{can only be entered once} and also \emph{dominates execution time} as
the \emph{\mainloop}. \system applies this algorithm to each thread of
execution independently, which enables automatic identification of all serving
phases in applications that use different types of work threads (\eg
multiple services).
This can mitigate attacks where a compromised serving thread uses another one
as a confused deputy~\cite{deputy} to perform syscalls on its behalf.

\item
In order to accurately determine the system calls required during the serving
phase, it is necessary to calculate the code partition that is accessible from
the beginning of the \mainloop. To accomplish this, \system utilizes static
analysis of the server binary and all of its library dependencies to construct
safe, albeit conservative, versions of the program's function-call graph (FCG)
and control-flow graph (CFG). These graphs are then used to compute the
serving partition and the system calls made from it.
Although reverse-engineering arbitrary binary software can be challenging,
recent research~\cite{sysfilter, egalito} has demonstrated that it can be
accomplished in a sound manner for modern x86-64/ARM Linux binaries. Moreover,
\system introduces a combination of value-flow analysis (VFA) and heuristics,
among other static analyses, to increase the precision of the FCG and to
resolve the names of libraries and functions loaded at run time (\eg via
\code{dlopen()} and \code{dlsym()}).
\end{points}

\begin{figure*}[ht!]
	\centering
	\includegraphics[width=\linewidth]{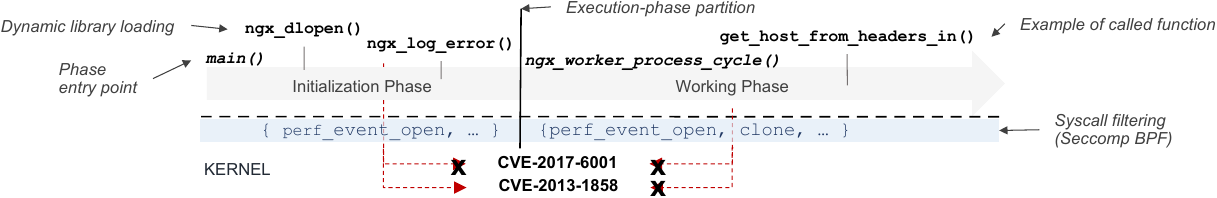}
	\caption{\emph{Motivating example}--System call filtering prevents
		compromised processes from performing system calls to interact with the OS
		or exploit kernel vulnerabilities. Installing more restrictive filters for
		the working phase of processes, like the \nginx web server's serving
		phase, further eliminates attacks and limits attacker capabilities.
		Existing approaches depend on manual defining the execution-phase
		partition, ignore dynamic library loading, and require source code and
		recompilation.}
	\label{fig:motivation}
\end{figure*}

\paragraph{Evaluation:}

\system was implemented for x86-64 Linux, using static and dynamic analyses
built as passes over the Egalito framework~\cite{egalito} and run-time tools
over Intel's Pin framework~\cite{pin:pldi05}, respectively. The FCG was further pruned
using TypeArmor~\cite{typearmor}, and application binaries were rewritten
using Egalito to install a Linux \seccomp filter just before the \mainloop.
We evaluated \system using six popular servers (\nginx, Apache \httpd,
\lighttpd, \bind, \memcached, and \redis), demonstrating that it outperforms
\sf and performs closely to \tsp, even without source code access. On average,
\system only allows 8.33\% more syscalls than \tsp, and filters as many
security-critical syscalls as \tsp in 88.23\% of cases. In fact, \system
outperforms \tsp in 2\% of cases.
\system is effective in thwarting exploit payloads, with a success rate of
53.83\% - 78.5\% and a success rate of 36.11\% - 77.77\% in blocking kernel
vulnerabilities, for the ones tested, depending on the application. When
considering libraries loaded at
runtime, \system outperforms \sf in resolving instances of \code{dlopen()} and
\code{dlsym()} by 58.33\% and 18.75\%, respectively, and resolves all
related calls in \redis server entirely with static analysis. We also evaluated \system on Abyss Web Server~\cite{Abyss}, which is a closed-source server. Last, on average, the static analysis component of \system runs 80\% faster than \tsp.

\paragraph{Contributions:}

Our main contributions are as follows:

\begin{itemize}

\item We introduce \system, a system that can significantly limit the number
of available syscalls during the serving phase of a server application, even
without access to the application's source code.

\item We design a novel algorithm for automatically detecting the \mainloop
corresponding to the serving phases of a server application's threads and
processes.

\item We propose a novel static analysis that combines VFA, TypeArmor, and
heuristics to refine the FCG and resolve dynamically-loaded libraries and
functions.

\item We evaluate \system using six server applications.
	\begin{itemize}
	\item In terms of security benefits, \system surpasses the state-of-the-art
	binary-only solution \sf by tightly restricting the available syscalls during
	the serving phase of server applications. It also performs closely to \tsp (the
	state-of-the-art source-code solution).

	\item \system beats \sf in resolving DLL-related calls and, to the best of
	our knowledge, is the first to soundly handle a server application loading
	libraries at run time.
	\end{itemize}

\item We make \system available at \url{https://github.com/vidyalakshmir/SysPartArtifact.git}.

\end{itemize}

\section{THREAT MODEL}
\label{sec:threat}

Binary server applications may contain
vulnerabilities in the application or the used libraries, which can be
exploited~\cite{szekeres2013sok} during the serving of requests. These
vulnerabilities can allow attackers to execute arbitrary code using techniques
like code injection and code reuse \cite{sd_ret2libc, shacham2007geometry,
	younan2012runtime}, bypassing popular defenses
such as stack-canaries~\cite{stackguard}, ASLR~\cite{aslr}, DEP~\cite{dep},
and others~\cite{cfi:ccs05} using known methods \cite{schuster2015counterfeit,
	outofcontrol:oakland14, threadspraying:usenixsec16}.

This work focuses on what a compromised process can do after this point. If
the process is unprivileged, attackers frequently exploit a vulnerability reachable through a system call to elevate their privileges
\cite{kemerlis2012kguard, kemerlis2014ret2dir, pomonis2017kr,
	pomonis2019kernel, kemerlis2015protecting, li2017lock}. Compromised
programs, also use system calls to perform malicious actions, which are part
of their payload, such as downloading and executing malicious software,
attacking other servers over the network, and more. In multi-threaded/process
applications, where the compromised thread is restricted and cannot perform a
vulnerable system call, attackers may attempt to confuse another
(unrestricted) thread or process into performing the syscall on their behalf
\cite{deputy}. This can be achieved, for example, by corrupting the data used
by the other thread or sending malformed data to another process through
inter-process communication (IPC) channels.

\section{BACKGROUND AND MOTIVATION}
\label{sec:background}

\subsection{Filtering System Calls using \seccomp}

Filtering the unused system calls of a process is one way to limit the attack
surface of the kernel and what a process can do in adherence to the principle
of least privilege. \seccomp~\cite{seccomp} is a Linux kernel facility which
allows filtering of system calls using Berkeley Packet filter rules. The
process of filtering system calls using \seccomp is one-way, which means the
filtered system calls cannot be re-allowed later as this would open a window
for attackers to reactivate them once an application is compromised. \seccomp
filters are manually defined which is an error-prone process as determining
the system calls that are actually needed by an entire application is often
complex.

\subsection{Automating Filter Generation in Binaries}

Previous works, such as \sf~\cite{sysfilter}, aim to automate the process of
installing \seccomp filters on binaries. This is achieved by disabling
syscalls that are not required during the binary program's lifetime. Using
static analysis, \sf first constructs the function call graph of the binary
and extracts the syscalls reachable from its code. It then injects code in the
binary to install the generated \seccomp filters at load time.

\subsection{Temporal System-Call Filtering}
\label{sec:tspback}

Compared to limiting syscalls for the whole execution, a more effective
approach is temporal system-call filtering: \textit{installing progressively
	more restrictive system-call filters as an application executes}. For example,
most server programs work in two phases. \ding{182} The
\textbf{initialization} phase performs tasks like initializing configuration
parameters, forking service processes, spawning worker threads, \etc
\ding{183} Upon completion of the initialization phase, it enters the
\textbf{serving} phase, which is usually designed as a loop---or the
\textbf{\mainloop}--- to continuously handle client requests. We define the
point of transition from the initialization phase to the serving phase as the
\textbf{transition point}. Many syscalls are no longer needed after this
transition point, which can be filtered out to further constraint the
application.

\paragraph{Running Example:}

In \autoref{fig:motivation}, we show a running example on Nginx to demonstrate
the security benefits of temporal system-call filtering. The serving phase of
\nginx starts at the function \texttt{ngx\_worker\_process\_cycle()}. By
exploiting vulnerabilities like CVE-2013-2028~\cite{nginxbug}, adversaries can
control the execution of the serving phase through ROP. Temporal syscall
filtering can help prevent such adversaries from moving deeper by elevating
the privileges of the controlled process.
For instance, it can filter out the \texttt{clone} syscall during the serving
phase, disallowing the adversaries to compromise the kernel via vulnerability
CVE-2013-1858~\cite{clonebug}. In contrast, whole-binary syscall filtering
like \sf cannot achieve this as \texttt{clone} is needed in the initialization
phase.

\paragraph{Existing Solutions:}

\tsp~\cite{temporal} is the state-of-the-art solution for temporal syscall filtering.
\tsp runs semi-automatic analyses on the source code to determine the serving phase and the required syscalls. It further
mounts a syscall filter at the transition point to disable the unneeded syscalls. While insightful, the design of \tsp
has several limitations that can restrict its practicality.

\begin{points}
    \item  \tsp requires source code.   \textcolor{\revcolor}{It cannot work on closed-source
	applications that are common in various domains: (i) Commercial servers like Aprelium Abyss Web Server, SAP NetWeaver, Oracle Database, etc., are only provided in the format of binary; (ii) Many closed-source desktop applications, such as Zoom and Skype, act as both clients and servers; (iii) Governments often acquire and run specialized proprietary software, including Linux servers, as attested by their interest in securing them~\cite{tpcp}; (iv) Cloud users can run binary servers without sharing the source code for cloud vendors to provide protection. Even for open-source applications, gathering the correct version of source code and all its dependent libraries is cumbersome and
	can be infeasible (e.g., open-source servers can use proprietary binary libraries). In addition, inline assembly can arise in source code (like those in GNU libc) and fail source-code-based approaches.}

    \item \tsp requires users to identify the transition point manually. This
	is more complicated and time-consuming than it appears to be. For
	instance, \memcached offers a diverse set of services (client request
	handling, LRU maintenance, slab rebalancing, \etc) via different worker
	threads. These services use different serving code and have different
	transition points. It is important to identify all of them and place a
	syscall filter on each of them. Otherwise, confused deputy attacks are
	possible: a compromised but filter-restricted thread can hijack another
	unrestricted thread to invoke an unavailable syscall. The manual effort
	needed and the possibility of errors dramatically increase in such cases.

	\item \tsp currently does not handle dynamically-loaded libraries, which can
	cause critical functionality issues. For instance, \httpd may use
	\texttt{mod\_ssl.so} to enable SSL and TLS connections, requiring 14
	additional system calls. Handling dynamically-loaded libraries presents a
	non-trivial challenge for server applications, which often follow a
	modularized design and load these libraries at run time using interfaces like
	\texttt{dlopen()} and \texttt{dlsym()} based on a configuration file. However,
	resolving the libraries and functions imported by an application through
	these interfaces is necessary to avoid errors.
\end{points}

\begin{figure*}[t]
	\centering
	\includegraphics[width=\linewidth]{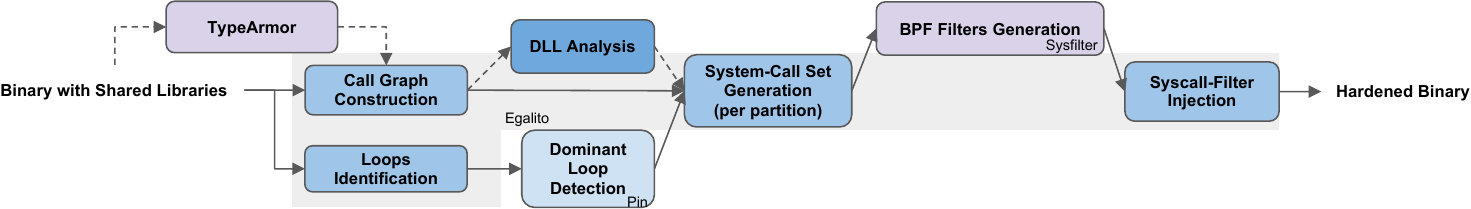}
	\caption{Workflow of \system.}
	\label{fig:sysdesign}
\end{figure*}

\section{DESIGN AND IMPLEMENTATION}
\label{sec:design}

\subsection{Overview}

The objective of \system is to overcome the limitations of previous
works~\cite{sysfilter, temporal} and generate an automatic system-call filter
for server binaries. To achieve this, our approach uses both static and
dynamic analyses, as illustrated in \autoref{fig:sysdesign}. All analyses
operate on binary code. Static analyses are based on the
Egalito~\cite{egalito} framework, while dynamic analyses use Intel's
Pin~\cite{pin:pldi05} framework. To automatically identify the serving phase
of server binaries, we first use static analysis to locate loops, and then employ
dynamic analysis to determine the dominant loop for each thread, which
corresponds to its \mainloop. Before generating system-call filters, we
construct the function call graph of the application statically, and then
refine it using value-flow analyses (VFA) and heuristics. We also use
TypeArmor~\cite{typearmor}, a third-party tool, to further refine the graph.
To determine the use of dynamically-loaded libraries (DLL) by the application,
we combine both static and dynamic analysis, with the former providing
soundness. The above information is then combined to generate a list of
required system calls for each serving phase. Next, \system generates a BPF program that will install the appropriate \seccomp filters, which is inspired from \sf. Finally, using Egalito, we rewrite the application and inject these functions at the phase transition
points. All extensions that we developed over Egalito is described in section ~\ref{sec:a5}.

\subsection{Serving Phase Detection}
\label{sec:findpartition}

To define the beginning of the serving phase of a server program, we build on
the following observations:
\begin{itemize}
\item Server applications utilize a loop to serve clients.

\item A system-call filter should be installed outside a loop to avoid
repeated installation.
\end{itemize}

Hence, we define the serving phase of a program as the beginning of a
top-level loop, where it spends most time executing. Here, top-level means
the loop is not enclosed in another loop. Our algorithm for detecting the
serving phase, therefore, focuses on identifying this \emph{\mainloop}.
Preceding it is the transition point, where a system-call filter can be
installed. Serving phase detection consists of a static-analysis phase (\S\ref{sec:staticloop}),
which finds all loops in the application and its libraries, and a
dynamic-analysis phase (\S\ref{sec:dynamicloop}), where this information is used
to find the dominant or \mainloop. By design, this algorithm can find the
\mainloop used by different worker threads (or processes) of the server,
enabling \system to generate system-call filters for all of them. We describe
them below.

\subsubsection{Loops Identification:}
\label{sec:staticloop}

To identify all loops, we employ the Egalito~\cite{egalito} binary-analysis
tool to statically disassemble applications and libraries, and extract the
control-flow graph (CFG) of each function of the application and its libraries. We focus on loops
produced using: \code{for() \{\}}, \code{while() \{\}}, \code{do\{\} while()},
and \code{goto} statements. For each function, we employ the worklist
algorithm~\cite{Loops, looplowry, loopsutter}, which builds on the concept of
dominance, to identify all loops present in the function. Briefly, the
algorithm is based upon finding \emph{dominator} nodes and \emph{back edges},
as defined below.

\paragraph{Dominator:}

A node N (which represents a basic block in a CFG) dominates node M if all
paths to M must pass through N. A node dominates itself. If there is an edge
from node N to M in the CFG, then N is a predecessor of M and M is a successor of N. The
set of dominators of node Z is defined as:

$dom(Z) = {Z} \cup  (\cap (dom(Y))$,

where $Y$ = set of all predecessors of $Z$.

\paragraph{Back Edge:}

An edge from node N to node H is a back edge if H dominates N. Node H
is the ``header'' of the loop (the place where the loop is entered).

Using the above, the algorithm identifies the code blocks comprising the
\emph{body} of loops, as well as the \emph{exit nodes} of each loop:

\paragraph{Loop Body:}

For a loop with back edge from node N to node H (header node), the body of the
loop is obtained by starting from node N and recursively finding all
predecessor nodes until the header node is reached.

\paragraph{Exit Nodes:}

The exit nodes of a loop are determined by identifying those nodes within the
loop body who have outgoing edges to nodes outside of the loop.

\begin{algorithm}
\caption{Dynamically compute the executed top-level loops}
\label{alg:domloop}
\scriptsize

\begin{algorithmic}[1]
\Require A set of loops \code{S} each with a \code{start\_address} and a set of exit\_address
\Ensure Set \code{executed\_top\_loops} which contains executed top-level loops
\Function{compute\_outer\_loop}{\code{S}}
\State \code{cur{\_}loop} $\gets$ []
\For{Each instruction with address \code{i}}
	\For{Each loop \code{L} in \code{S} whose \code{exit{\_}address} is \code{i}}
		\If{\code{cur{\_}loop} is \code{L}}
			\State \code{endtime(cur{\_}loop)} $\gets$ \code{get\_current\_time()}
			\State \code{duration(cur{\_}loop)} $\gets$ \code{endtime(cur{\_}loop)} - \code{starttime(cur{\_}loop)}
			\State \code{executed\_top\_loops} $\gets$ \code{executed\_top\_loops} $\cup$ \code{cur{\_}loop}
			\State \code{cur{\_}loop} $\gets$ NULL
			\State \code{break}
		\EndIf
	\EndFor
	\If{\code{i} is the entry{\_}address of loop \code{L}}
		\If{\code{cur{\_}loop} is NULL}
			\State \code{cur{\_}loop} $\gets$ \code{L}
			\State \code{starttime(L)} $\gets$ \code{get\_current\_time()}
			\State \code{iterations(L)} $\gets$ 0

		\ElsIf{\code{cur{\_}loop} is \code{L}}
			\State \code{iterations(L)} $\gets$ \code{iterations(L)} + 1
		\EndIf
	\EndIf
\EndFor
\EndFunction
\end{algorithmic}
\end{algorithm}

\subsubsection{Dominant Loop Detection:}
\label{sec:dynamicloop}

We implement a component using Intel's Pin~\cite{pin:pldi05} to determine the dominant loop when running servers with simple workloads . Intel Pin is a dynamic binary instrumentation tool for developing and applying tools, known as pintools, on
binary programs at run-time. We develop a pintool that utilizes the data
obtained statically in the previous step, including the start address of the
loop and the addresses of its exit nodes, to calculate the amount of time each
top-level loop encompasses execution for each process and thread, according to
Algorithm~\ref{alg:domloop}.

The tool takes a set of loops S as input, where each loop has a starting
address and a set of exit addresses. The algorithm iterates through each executed
instruction in the program and, for each instruction, checks whether it is the
entry address of a loop in S. If it is, and if the currently executing loop (\code{cur\_loop}) is NULL, then it sets \code{cur\_loop} to this loop and sets its start time and iteration count to zero. 
If \code{cur\_loop} is not NULL,
the algorithm increments the iteration count for \code{cur\_loop}. If the instruction
is an exit address of \code{cur\_loop}, the algorithm calculates the
duration of the loop  and adds it to the output set of executed loops. This
process is repeated for each instruction in the program until the program
exits. At the end of the algorithm, the set of executed loops contains all the
top-level loops that were executed, along with their start time, end time,
duration, and iteration count. The information is organized per
thread/process. We consider the loop with the largest execution time as the
\mainloop of each execution thread, and its start address the transition point
where filter installation should be placed.

\subsection{Call Graph Construction}

In order to compute the system calls needed by each \mainloop, we need to
determine the code that is reachable after the transition point. As a first
step, we generate the function-call graph (FCG) of the program rooted at
\code{main()} and across shared libraries. Although, we do not include any
code running before \code{main()}, such as initialization routines, we do
account for how it may affect the FCG (discussed in \S\ref{sec:at}). The
method described here is based on Egalito's VacuumFCG pass, which is also used
by \sf and demonstrated to be sound. Below, we summarize the methodology.

\subsubsection{Computing Direct Edges:}

FCG construction starts from the functions that the Linux program loader will
call, which includes initialization functions in \code{.init},
\code{.init\_array}, and \code{.preinit\_array}, cleanup functions in
\code{.fini} and \code{.fini{\_}array}, and  the program's \code{main()}.
Egalito disassembles each function's instructions and follows direct branches,
such as calls, jumps, and conditional branches, to construct an initial CFG.
When there are calls to functions in shared libraries, those edges initially
point to the binary's procedure linkage table (PLT). Egalito emulates the
dynamic linker/loader to resolve them and point them to the actual function,
extending (this way) the graph across shared libraries. Egalito is robust
enough to also handle GNU libc. Also, we resolve calls to functions over the Name
Service Switch (NSS) scheme using the approach mentioned in \sf.

\subsubsection{Computing Indirect Edges:}
\label{sec:at}

The FCG includes indirect control-flow edges that correspond to calls made
through function pointers, which are represented by indirect call or jump
instructions in the basic blocks of the disassembled binaries. However, it is
difficult to resolve the targets of these indirect branches
\cite{ramalingam1994undecidability}. Therefore, \system over-approximates the
set of functions that these branches can target. Specifically, it includes all
address taken (AT) functions, which are functions whose start address is
referenced or loaded in the program either directly or by taking the address
of a PLT entry. To identify AT functions, VacuumFCG takes advantage of the
fact that modern Linux binaries are built as position-independent code (PIC)
and include metadata in the form of relocation entries for every function
address taken in the program \cite{dinesh2020retrowrite}.

Many function pointers in programs and libraries are part of the
initialization of constant function-pointer arrays. To reduce the number of AT
functions, VacuumFCG prunes~\cite{nibbler:acsac19,libfilter:dtrap20} those
that are included in such arrays but are never used or taken by live (\ie
actually used) program code. When symbols are available, the analysis can
calculate the boundaries of data objects using symbol information and consider
only the AT functions that are in objects used by the application. This
process is performed iteratively to find objects and AT functions actually
used by the program. If symbols are not available, any program address
pointing to a section of the binary leads to the inclusion of all AT functions
found in it.

\subsection{Refining the FCG}
\label{sec:refining}

Prior works~\cite{nibbler:acsac19,libfilter:dtrap20,sysfilter} operating on
binaries had to rely on the above method of over-approximating the FCG, by
assuming that all indirect calls can target any AT function, to ensure
soundness. To increase the precision of the FCG, \system implements data-flow analysis for constants (\ie value-flow analysis or VFA),
and incorporates \typearmor, a binary-analysis system that aims to reduce the
number of possible targets for indirect call sites.

\subsubsection{Forward Value-Flow Analysis:}
\label{sec:pruneatvfa}

\system employs forward VFA to determine where each function pointer flows and
eliminate it, if possible, from the list of AT functions obtained in
\S~\ref{sec:at}. If the pointer is only used as the target of an indirect call, it is removed from the list of AT functions and  an edge between the indirect
call and pointed function is also added to the FCG.
If the pointer is not passed as an argument to a function, nor stored in memory nor returned as the value of a function, then it is removed from
the list of AT functions. For a function to be completely eliminated from the list, all of its pointers need to be resolved in this manner.

The example in \autoref{lst:nginxvfa} shows such a case from
\nginx, where a pointer to the
\code{ngx\_http\_upstream\_init\_round\_robin()} (located at
\code{0x7e1d(\%rip)}) is taken on line 6 and is later used by the indirect
call in line 3. The only use of the function pointer is as target of the indirect call and hence can be removed from the AT list.

\begin{lstlisting}[style=x86,label=lst:nginxvfa,caption={Flow of function pointer in the \nginx server.}]
bbl2:
    movq         %r12, %rdi
    call         *%rax
    ...
bbl1:
    leaq         0x7e1d(%rip), %rax  ; <ngx_http_upstream_init_round_robin>
    jmp		     0x6b2d2              ; <bbl2>
\end{lstlisting}

Likewise, the pointer to \code{RedisModuleCommandDispatcher()} in
\autoref{lst:redisvfa} is only used in a compare instruction in line 3.
We can safely remove it from the AT list, because it cannot be the target of
any indirect call.

\begin{lstlisting}[style=x86,label=lst:redisvfa,caption={Flow of function pointer in the \redis server.}]
bbl1:
    leaq         -0x21c9(%rip), %rbp       <RedisModuleCommandDispatcher>
    cmpq         %rbp, 8(%rax)
    (JUMP jne)   0xb1948
\end{lstlisting}

We implement VFA using Egalito's use-def chains.
Use-def analysis, tracks the uses and definitions of a register or memory location. A register or memory location is defined when a value is written to it, and it is considered used whenever its value is read.  The use-def chains, provided by Egalito, also maintain the locations in successor basic blocks where a register or memory location is
later used, and the locations in predecessor blocks where it was previously
defined.
For example in \autoref{lst:usedef}, use-def analysis determines that the instruction at address \code{0x0002a7aa}, defines register \code{rdx} and uses
\code{rcx}. Use-def chains also inform us that \code{rcx} was previously
defined by the instruction at \code{0x0002a72c} and \code{rdx} is later used by
the instruction at \code{0x0002a7b1}.
Even though Egalito does not maintain use-def chains across function calls,
our VFA is inter-procedural as it tracks values passed (forward) to functions
through registers (\eg as arguments).

\begin{lstlisting}[style=x86,label=lst:usedef,caption={Use-def information provided by Egalito.}]
0x0002a72c:         movl         8(%rdi), %rcx
0x0002a7aa:         movl         %rcx, %rdx
0x0002a7b1:         addl         %rdx, %rdi
\end{lstlisting}

\subsubsection{Backward Value-Flow Analysis:}

\system leverages use-def chains to perform backward value-flow analysis starting from indirect calls to determine the possible values of the operand of the indirect call instruction. If a path points to a value that is a function pointer, an edge to it is added from the function containing the indirect call. If all paths lead to values from function pointers, the indirect call is no longer assumed to target all the functions in the AT list, considerably improving the precision of the FCG. However, even if one path leads to a memory
load the analysis for that indirect call terminates. This analysis is also partially inter-procedural, as the forward one.

The example in \autoref{lst:nginxback} shows such a case, where two functions, lines 1 and 5,
call \code{ngx\_sort}, passing a function pointer as an argument, through the
\code{rcx} register. This call now has exactly two possible targets that it
can call into. Our backward VFA starts from line 11 to determine the value of register \code{r15}. Register \code{r15} is referenced in line 10 and its value is defined to be loaded from \code{rcx}. Since \code{rcx} is an argument register, inter-procedural analysis is performed to find all invokations of \code{ngx\_sort()}. From the usedef information at the call sites of \code{ngx\_sort()} (lines 3 and 7), the analysis determines that register \code{rcx} is referenced at line 2 and line 6, respectively. The analysis terminates at lines 2 and 6, where a function address is loaded into \code{rcx}. Hence, the analysis is able to completely resolve the indirect call target at line 11.

\begin{lstlisting}[style=x86,label=lst:nginxback,caption={Resolving all possible values of an indirect call in the \nginx server.}]
ngx_resolver_process_response:
  leaq     -0x59a5(%rip), %rcx     ; <ngx_resolver_cmp_srvs>
  call     0x25a33                 ; <ngx_sort>
  ...
ngx_http_block/bb+5407:
  leaq     -0x1d43(%rip), %rcx     ; <ngx_http_cmp_conf_addrs>
  call     0x25a33                 ; <ngx_sort>
  ...
ngx_sort:
  movq     %rcx, %r15
  (CALL*)  *%r15
  ...
\end{lstlisting}

\subsubsection{TypeArmor:}
\label{sec:typearmor}

\system uses TypeArmor~\cite{typearmor} to further prune the number of targets
for each indirect-call site. TypeArmor is a static analysis tool for binaries
that aims to refine the set of functions that can be targeted by indirect
function calls. It does so by detecting the signature of call sites and
functions. For each indirect call, it calculates the number of arguments
prepared and whether it expects a return value. For each function, it
calculates the maximum number of arguments it expects and whether it returns a
value. Call sites with $n$ arguments are matched to AT functions that expect
$\eqslantless n$ arguments and similar return behaviors (value \vs no value
returned).

\subsection{Dynamically-Loaded Libraries Analysis}
\label{sec:dynamic}

In many servers, optional, non-core functionality is sometimes activated at
run time, depending on the presence of a library on the system or
configuration options. When such functionality requires additional libraries,
those are loaded using \code{dlopen()}, which loads the library
with \code{filename} into the address space. Interfaces to DLLs are obtained
through \code{dlsym()}, that takes as input the handle returned by \code{dlopen()} and the
name of a symbol, which could be an exported function or global variable.
Function pointers returned by \code{dlsym()} can be called through an indirect
call and may perform system calls, hence, it is necessary to resolve the
libraries loaded and symbols queried through these two functions.

\system employs static analysis to recover this information. To handle cases
where the static analysis fails to resolve all possible values, we use
dynamic training by running the application with a desired configuration and
common workloads.

\subsubsection{Static Analysis:}

\system uses two approaches to statically discover this information: backward
VFA and a heuristic.

\paragraph{Backward VFA:}

We leverage this type of analysis again to find the \code{filename} and
\code{symbol} arguments used in \code{dlopen()} and \code{dlsym()},
respectively. If the application is using constant strings, we can find
pointers to those strings flowing to the first and second argument,
respectively. In x86-64 Linux, these correspond to registers \code{rdi} and
\code{rsi}.

\paragraph{Heuristic:}

The interfaces queried using \code{dlsym()} are frequently hard-coded in
applications, while the names of the libraries to be loaded are provided in
configuration. For example, the \bind server includes a plugin for accessing
Samba Active Directory (AD) databases. The interface to this plugin includes a
set of functions (\code{dlz\_*}) which are constant. However, depending on the
AD database in use, the user can load a different plugin version with a
configuration like the following.

\begin{lstlisting}[style=Cstyle, numbers=none, caption={Bind configuration file specifying plugin.}]
dlz "AD DNS Zone" {
    # For BIND 9.16.x
    # database "dlopen /usr/local/samba/lib/bind9/dlz_bind9_16.so";
};
\end{lstlisting}

Based on this observation, when we are able to resolve all possible values
flowing into \code{dlsym()} call sites, but not to \code{dlopen()}, we use
the first to search the system for libraries exporting any of resolved
symbols. We consider all matching libraries as potential inputs to
\code{dlopen()} and include them in our analysis.

\subsubsection{Dynamic Analysis:}

To discover the libraries loaded at run time, we run applications with desired
configuration options and common workloads, and intercept calls to
\code{dlopen()} and \code{dlsym()} to record their arguments. We use function
interposition to redirect calls to our functions, which record their arguments
and proceed to call the original functions. This is done by creating a shared
library defining these two functions and pre-loading it through the
\code{LD\_PRELOAD} environment variable on Linux, when launching the
application.

\subsubsection{Incorporating Results:}

\system considers any libraries found in this step as additional dependencies
and resolved symbols are marked as taken at the call site of the corresponding
\code{dlsym()}. Finally, the static analysis described in
\S\ref{sec:pruneatvfa} is reapplied to again prune the list of AT functions
and indirect-call targets.

{\color{\revcolor}

\subsection{Handling \texttt{execve}}

The \texttt{execve} syscall allows a process to load and execute a new program,
so it needs special handling. Similar to how we deal with DLLs, \system combines
static analysis and dynamic analysis to collect the arguments passed to \texttt{execve}---in
particular the path of the new program to be executed. First, \system runs static, backward VFA to
find the arguments of \texttt{execve}. Second, \system traces the execution of the application
under desired configurations and common workloads to learn the arguments passed to \texttt{execve},
with the help of Pin. \system also offers users the option to add the list of programs
that can be executed through \texttt{execve}.

Based on the above analyses and user input, any new programs that may be launched
by \texttt{execve} during the serving phase  will be further analyzed to gather the syscalls
they require. The analysis results can be applied in two ways. First, the newly identified
syscalls will be added to the allowed list and the entire list will be propagated to
the new program. Second, we start with the extended allowed list for the initial program
but further reduce the list every time \code{execve} is invoked to only keep those syscalls
needed by the new program. This approach is inspired
by \sf~\cite{sysfilter}. %
}

\subsection{System-Call Set Generation}

Computing the complete set of system calls reachable from the transition point
is crucial to determine the correct system-call filter for the serving phase, as
erroneous filters can lead to program termination. In order to
determine the system calls of each serving phase, first we compute the system
calls which are invoked by each function within the FCG. Next, the code which
is reachable from the transition point of that serving phase is determined.
Finally, we collect all system calls which are invoked from the reachable
code.

\subsubsection{Finding System Calls Reachable from Each Function of the FCG:}
\label{sec:sysfunc}

A system call is represented by a system-call number and may or may not have
arguments. Applications can invoke system calls by invoking libc
wrapper functions, which in turn invoke the system call (for example, \code{open()}
invokes the corresponding \code{open} syscall), use the \code{syscall()} libc
function, or use inline assembly and the \code{syscall} instruction.

\system uses Egalito's \code{FindSyscalls()} which is run on all the functions
in the server and its dependent libraries to find all system calls that are invoked directly from these functions. The pass parses each
instruction within a function and searches for \code{syscall} instructions. If
one is found, it employs backward VFA to determine the value of the register
\code{rax} that contains the system-call number to be executed. Similarly,
backward VFA is used for calls to \code{syscall()} to find the value of the
argument specifying the syscall number (\eg the register \code{rdi} in x86-64
Linux systems).

\begin{algorithm}
\caption{An algorithm to find all system calls reachable from all functions in \code{FCG} rooted at \code{main()}}
\label{alg:FINDSC}
\scriptsize
\begin{flushleft}
\textbf{Input}: \code{FCG}; Map \code{M(S, L)} where \code{L} is a list of functions which directly invoke syscall \code{S} \\
\textbf{Output}: Map \code{fsyscalls(F, L)} where \code{L} is the list of syscalls reachable from function \code{F}
\end{flushleft}
\begin{algorithmic}[1]
\Function{syscalls\_function}{\code{FCG,S}}
\For{Each \code{(s, flist)} in \code{M}}

    \For{Each \code{f} in \code{flist}}
        \State Push \code{(f,f)} onto stack \code{S}
        \State \code{processed} $\gets$ []
        \While{stack \code{S} is not empty}
            \State \code{(cur,cur\_child)} $\gets$ top of stack \code{S}
            \State Pop from stack \code{S}
            \If{\code{cur} is in \code{processed}}
                \State continue
            \EndIf
            \State \code{processed} $\gets$ \code{processed} $\cup$ \code{cur}
            \State \code{pp} $\gets$ Parents of \code{cur} in \code{FCG}
            \For{Each \code{parent} in \code{pp}}
                \State Push \code{(parent,cur)} onto stack \code{S}
            \EndFor
            \If{\code{cur} == \code{cur\_child}}
                \State continue
            \EndIf
            \State Insert \code{s} into \code{fsyscalls[cur]}
        \EndWhile
    \EndFor
\EndFor
\Return \code{func\_syscalls}
\EndFunction
\end{algorithmic}
\end{algorithm}

The set of system calls \textbf{reachable} from each function is computed
using algorithm~\ref{alg:FINDSC}, which works by combining the set of system
calls directly invoked by the function with those that are reachable from its
child functions in the FCG. It is a graph traversal algorithm that begins at
each function that invokes system calls directly, and iteratively propagates
the value of the system call to its parents.

\begin{algorithm}
\caption{An Algorithm to find system calls reachable from the transition point located at address \code{entry\_addr} within  function \code{f}}
\label{alg:SCPARTITION}
\scriptsize
\begin{flushleft}
\textbf{Input}: Address \code{entry\_addr}, Function \code{f}, \code{FCG} rooted at main() \\
\textbf{Output}: Set \code{SC}, which contains system calls reachable from \code{entry\_addr} \\
\end{flushleft}
\begin{algorithmic}[1]

\Function{syscalls\_partition}{\code{entry\_addr,f,FCG}}
\State \code{noreturnFns} = Find all noreturn functions
\State \code{threadFns} = Find all thread start functions
\State Push \code{(entry\_addr, f)} to stack \code{S}
\State \code{SC} $\gets$ syscalls\_of\_fini\_fns()
\While{stack \code{S} is not empty}
    \State \code{(addr,fun)} $\gets$ top of stack \code{S}
    
    \State Pop from stack $S$

    \State \code{result} $\gets$ syscalls\_invoked\_at\_instruction\code{(addr, fun)}
    \State \code{CFG} $\gets$ CFG of \code{fun}
    \State \code{visited} $\gets$ []
    \State \code{B} $\gets$ basic block at address \code{addr}
    \State Insert \code{B} into \code{visited}
    \For{Each successor \code{succ} of \code{B} in \code{CFG}}
        \State Push \code{succ} to stack \code{SS}
    \EndFor
    \While{Stack \code{SS} is not empty}
        \State \code{T} $\gets$ top of stack \code{SS}
        \State Pop from stack \code{SS}
        \State Insert \code{T} into \code{visited}
        \For{Each instruction \code{i} with address \code{iaddr} in \code{T}}
        	\State \code{cur} $\gets$ []
            \If{\code{i} is a call instruction}
                \For{\code{c} in call target set \code{CT} of \code{i}}
                    \State \code{cur} $\gets$ \code{cur} $\cup$ \code{reachable\_syscalls(c)}
                \EndFor
            \Else
                \State \code{cur} $\gets$ syscalls\_invoked\_at\_instruction\code{(iaddr, fun)}
            \EndIf
            \State \code{result}  $\gets$ \code{result} $\cup$ \code{cur}
        \EndFor
        \For{Each successor \code{succ} of \code{T}}
        	\If{\code{succ} is not in \code{visited}}
            	\State Push \code{succ} to stack \code{SS}
            \EndIf
        \EndFor
    \EndWhile

    \State \code{SC} $\gets$ \code{SC} $\cup$ \code{result}
    \If{\code{fun} is in \code{noreturnFns}}
    \State continue
    \EndIf
    \If{\code{fun} is in \code{threadFns}}
    \State continue
    \EndIf
    \State \code{P} $\gets$ Parents of \code{fun} in \code{FCG}
    \For{Each \code{p} in \code{P}}
        \State \code{calladdr} $\gets$ address at which \code{p} invokes \code{fun}
        \State Push \code{(calladdr, p)} to the stack \code{S}
    \EndFor
\EndWhile
\Return \code{SC}

\EndFunction
\end{algorithmic}
\end{algorithm}

\subsubsection{Finding System Calls Reachable from Transition Point:}
\label{sec:syscalltransition}

The transition point of the serving phase is the start address of the
\mainloop which can be represented by a tuple \code{(f, addr)}, where \code{f}
is the function containing the \mainloop and \code{addr} is the start address
of the \mainloop. Algorithm~\ref{alg:SCPARTITION} describes the algorithm which determines the code reachable from \code{(f, addr)} and the system calls invoked from the reachable code. The algorithm starts with non-returning functions analysis and computing thread start functions which are described below.

\paragraph{Non-returning Functions Analysis:}
\label{sub:sub:non-return}

We improved the no-return analysis pass in Egalito to find all non-returning
functions. A function \code{f} starting from basic block B is non-returning,
only if all paths from B in the \code{CFG} of \code{f} is non-returning.

\paragraph{Computing Thread Start Functions:}
\label{sub:sub:thread}

Threads are created by invoking the \code{pthread\_create()} function, with the thread start function specified as the third argument, which is the function executed as soon as the thread is spawned.
\system checks if
\code{pthread\_create()} function is a part of the FCG, and all invocations of
\code{pthread\_create()} are determined. Then, backward VFA
(\ref{sec:refining}) is employed at these callsites to find the third argument
passed to the function (register \code{rdx}).

\paragraph{Syscall-Set Computation:}
In order to find all system calls reachable from instruction \code{i} with address \code{addr} of function \code{f}, all code reachable from \code{i} has to be determined.
To start
with, we find all basic blocks within \code{f} which are reachable from
instruction \code{i}. Next, we consider all functions which are invoked from
within these basic blocks and get all system calls which are reachable from
all these functions (computed in section ~\ref{sec:sysfunc}). Next, if f
returns, then all callsites \code{(f', addr')} which invoke function
\code{f()} are found. Next, the analysis recursively proceeds to find all the
system calls which are reachable from \code{(f',addr')}. The algorithm stops
when main() is reached or a thread start function is reached or if function is
non-returning. Finally, it includes all system calls reachable from functions in .fini,.fini\_array and .dtors sections. In this way \system computes all code which are reachable from
the serving phase which also includes the code that the function returns back
to, which is not considered by \tsp.

\subsection{Enforcing System-Call Filters at the Partition Boundary}

{\color{\revcolor}
\subsubsection{Filter Policy:}
We enforce a policy that aims to only allow syscalls used in the application's main
loop. This policy has significant and immediate security benefits:
(i) attacking payloads are unable to utilize dangerous but blocked syscalls and
(ii) compromised applications cannot attack the operating system kernel using blocked syscalls. While further
restricting syscall arguments could additionally enhance security, it poses
challenges with Seccomp-BPF and is susceptible to TOCTOU
attacks~\cite{seccomp_toctou}. Note that previous
work~\cite{mishra2018shredder} has explored argument restriction for library
APIs.
}

\subsubsection{\seccomp Filter Generation:}

The list of allowed system calls of the serving phase obtained in
\S\ref{sec:syscalltransition} is used to create a C function which uses
Seccomp-BPF, as in \sf~\cite{sysfilter}, to install a filter that will be
enforced by the kernel. The generated function, \code{install\_filter()}, is
compiled into a shared library (\code{libsyspart.so}).

\subsubsection{Filter Insertion:}
\label{sub:sub:filter}

We developed a tool using Egalito that links against the generated library and
inserts call to \code{install\_filter} at the transition point and generates a
hardened binary. The server binary, the transition point represented by the
tuple \code{(f, addr)}, and \code{libsyspart.so} is fed as input to the tool.
The server binary and \code{libsyspart.so} are parsed using Egalito. The CFG of \code{f} is generated
and traversed to determine a basic block \code{B} such that B precedes
the basic block with address \code{addr} and B is not a part of the \mainloop. A new Egalito pass \code{SyspartPass} is then used to insert a function
call to \code{install{\_}filter()} after basic block \code{B}. A new hardened
binary with the installed filter is generated using Egalito \code{mirrorgen} output
generation mode.

\begin{table*}[ht]
	\centering
	\caption{Results of serving phase detection in \system compared with \tsp. 
	In \textcolor{red}{red}, we highlight the cases
	where \tsp demonstrably creates wrong filters. ``$\hookleftarrow$'' 
	indicates that the returning functions lead to errors in \tsp. 
	``Size $\Delta$'' stands for the number of more instructions 
	in \system's main loop than that in \tsp's serving-phase function.
	``\# of Other Loops'' is the number of extra loops detected by \system 
	for auxiliary server threads. ``Concurrency'' shows the number of processes (P)
	/ threads (T) used by each server application,  where ``*'' means one or more.}
	\vspace{-0.5em}
    \label{tab:loops}
	\begin{tabular}{l c | c | c c c}
		\toprule
        \multicolumn{2}{c}{\textbf{Application}} & \textbf{TSP} & \multicolumn{3}{c}{\textbf{\system}} \\
	    Name & Concurrency & Serving Function & Main Loop & Size $\Delta$ & \# of Other Loops \\
		\midrule
		\bind 		& 1/* & $\hookleftarrow$\badloopfunc{isc\_app\_ctxrun}& \loopfunc{main+0xe01} 						& +2.5K & 4 \\
        \httpd 		& */* & \loopfunc{child\_main}                        & \loopfunc{child\_main+0x598}   				& +0.3K & 5 \\
        \lighttpd 	& 1/1 & $\hookleftarrow$\loopfunc{server\_main\_loop} & \loopfunc{main+0x84} 						& +83K  & 0 \\
        \memcached 	& 1/* & \loopfunc{worker\_libevent} 				    & \loopfunc{event\_base\_loop+0xbb} 		& +1K   & 6 \\
        \nginx 		& */1 & \loopfunc{ngx\_worker\_process\_cycle} 	    & \loopfunc{ngx\_worker\_process\_cycle+0xbb}	& +0.3K & 1 \\
        \redis  	& 1/* & $\hookleftarrow$\loopfunc{aeMain} 		    & \loopfunc{aeMain+0x10} 						& +0.2K & 1 \\
		\bottomrule
	\end{tabular}
\end{table*}

\section{EVALUATION}
\label{sec:eval}

To evaluate \system and compare it with prior work, we apply it to the same
benchmark applications used by \tsp~\cite{temporal}: \bind 9.15.8 (incl. libuv-1.34.0 and
OpenSSL-1.1.1f), \httpd 2.4.39 (incl. apr-1.7.0 and apr-util-1.6.1), \lighttpd
1.4.54, \memcached 1.5.21 (incl. SASL-2.0.25 and libevent-2.1.11-stable) and
\nginx 1.17.1 (incl. OpenSSL-1.1.1f), and \redis 5.0.7.  All the dependent libraries
used by the applications are from Ubuntu 18.04.6 LTS, which is also adopted in the evaluation
of \tsp. We note that Ubuntu 18.04.6 uses glibc-2.27 by default while \tsp runs its static analysis
on glibc-2.24. To stay aligned, we also use glibc-2.24 for static analysis but glibc-2.27 for
dynamic library profiling (the older version cannot run). All the experiments
were performed on a 4-core Intel Core i7 8550U 1.80GHz CPU with 16GB of RAM,
running Ubuntu 18.04.6 LTS (kernel version 5.4.0-150).

\subsection{Serving Phase Detection}
\label{sec:srvphase}

{\color{\revcolor}

We first evaluate \system's ability to automatically detect the beginning of
the serving phase. To dynamically profile each server, we used its default
configuration. The algorithm is designed to identify the dominant top-level
loop in each thread, excluding the loops used in initialization and cleanup,
so even with some configuration or workload changes, the main loop remains the
same.

Simply launching the servers and letting them wait was sufficient to correctly
identify the serving phase with all servers, as it is common behavior for
servers to wait within their main loop. However, to ensure the \mainloop is
entered we assume that the desired workload is sent to the tested server. For
the evaluation, we used the following workloads: 10k HTTP requests for \httpd,
\lighttpd, and \nginx; 10k store/set requests of randomly generated key-value
pairs, followed by a retrieve/get request for each pair for \redis and
\memcached; and 10k IP address queries using the utility \texttt{dig} against
\bind.

In certain cases, the server can be set up to run in different modes,
which correspond to different \mainloop{s}. For instance, \nginx can be configured
as a proxy or cache server instead of a web server (the default configuration).
To capture the main loops of these different servers within \nginx, the user only needs to
profile it with the appropriate configuration. Once properly configured,
our approach automates the detection of different main loops, relieving the
user from the burden of code review. In our evaluation,
we did not test \nginx as a proxy or cache server.
}

Table~\ref{tab:loops} summarizes the results of the experiment. \system is
able to automatically detect the \mainloop corresponding to the serving
process/thread of each server, whose results are summarized in
\autoref{tab:allpartitions}. We further compared the main loops with
the serving-phase function manually identified by \tsp developers. In three
applications (\httpd, \nginx, \redis), \system identifies the \mainloop in the
\tsp's serving-phase function. In the case of \memcached, the \mainloop runs
inside a child function of \tsp's serving-phase function
(\code{worker\_libevent()}~\textrightarrow~\code{event\_base\_loop()}).
In the other two applications (\bind and \lighttpd), the situation is reversed.
\tsp's serving-phase function is invoked inside the \mainloop identified by \system.

\begin{figure}[ht!]
  \centering
  \includegraphics[width=\linewidth]{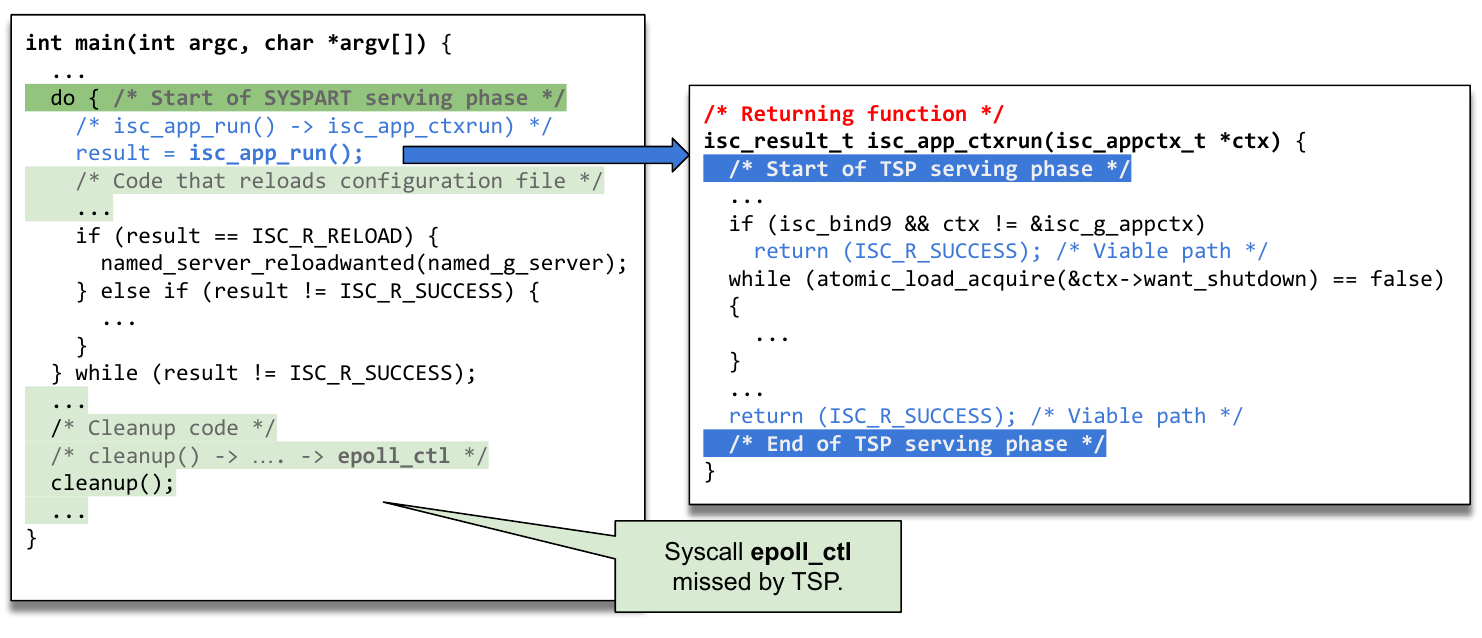}
  \caption{Advantage of \system's automatic detection of serving phase over
  \tsp's manual approach.}
  \label{fig:partitionboundary}
\end{figure}

Quantitatively, the size of the serving phase detected by \system is only
slightly larger than \tsp except for \lighttpd. The increase of size---including
the case of \lighttpd---is mainly because we consider the code following the return of the serving phase function (if it returns) and the finalization functions while \tsp does not consider these. Our choice is desired as ignoring
the post-serving code can omit system calls needed for
reloading configuration data or cleanup before termination.
This can potentially lead to issues like corrupting server data
and leaking operating systems resources. One such example from \bind
is shown in ~\autoref{fig:partitionboundary}. \tsp identifies
\code{isc\_app\_ctxrun()} as the serving-phase function. When it
returns, the cleanup code attempts to performs an \code{epoll\_ctl} system
call, which is filtered by \tsp since it doesn't include it within its serving phase and leads to abrupt server termination.

Another advantage of \system is its ability to identify the serving phase for
auxiliary server threads(Table ~\ref{tab:allpartitions}).  For five
applications, \system detects at least one auxiliary serving phase. In the
case of Memcached, the number of auxiliary serving phases increases to six.
This enables us to filter out a safe but also tight set of system calls for
each server thread.

\subsection{Filtered System Calls}
\label{sec:filtered}

\begin{table*}[ht]
\caption{Number of system calls allowed with \system (binary), compared to
	\tsp (source code) and \sf (binary). ``\tsp'' refers to the publicly available
	prototype, and the numbers in parentheses correspond to the results in the paper.
	\tspfixed stands for the version of \tsp with our fixes incorporated
	(the fixes used are indicated by the subscript on each application).
	The entries highlighted in \textcolor{red}{red} indicate a confirmed error
	occurred in \tsp's definition of the serving phase.\label{tab:syscall}}
\vspace{-0.75em}
\centering
\begin{tabular}{l | c | c c | c c | c c c c}
\toprule
\textbf{Application} & \textbf{\sf} & \multicolumn{2}{c|}{\textbf{\tsp}} & \multicolumn{2}{c|}{\textbf{\tspfixed}} & \multicolumn{4}{c}{\textbf{\system}} \\
 & All & \code{main()} & Serving & \code{main()} & Serving & \code{main()} & Serving & Main loop & AT Functions \\
\midrule
\bind\textsubscript{\ref{prob2}, \ref{prob1}, \ref{prob7}, \ref{prob8}, \ref{prob9}, \ref{prob10}, \ref{prob11}} & 119 & 100 (99) & \textcolor{red}{86 (85)} & 100 & \textcolor{red}{83} & 112 & 103 & 103 & 102 \\
\httpd\textsubscript{\ref{prob2}, \ref{prob3}, \ref{prob4}, \ref{prob8}, \ref{prob9}, \ref{prob11}} & 98 & 94 (94) & 79 (79) & 94 & 80 & 92 & 92 & 92 & {\color{\revcolor} 92}\\
\lighttpd\textsubscript{\ref{prob2}, \ref{prob3}, \ref{prob6}, \ref{prob8}, \ref{prob9}} & 99 & 95 (95) & 76 (76) & 95 & 76 & 95 & 80 & 93 & 74 \\
\memcached\textsubscript{\ref{prob2}, \ref{prob3}, \ref{prob4}, \ref{prob8}, \ref{prob9}, \ref{prob11}} & 104 & 99 (99) & 84 (84) & 100 & 85 & 98 & 82 & 85 & 82\\
\nginx\textsubscript{\ref{prob2}, \ref{prob1}, \ref{prob3}, \ref{prob5}, \ref{prob8}, \ref{prob9}, \ref{prob11}} & 115 & 106 (104) & 97 (97) & 109 & 100 & 108 & 104 & 104 & 104\\
\redis\textsubscript{\ref{prob2}, \ref{prob3}, \ref{prob8}, \ref{prob9}} & 104 & 90 (90) & 82 (82) & 91 & 83 & 92 & 85 & 85 & 85\\
\bottomrule
\end{tabular}
\end{table*}

In this experiment, we compare the system calls allowed by \system with the ones
generated by \tsp and \sf. As \tsp do not resolve dynamically-loaded
libraries, we ignore them in this experiment. We consider four
different code partitions. \ding{192} \code{main()} includes all the code after the
\code{main} function is entered. \ding{193} \emph{Serving} refers to \tsp's serving phase. \ding{194}
\emph{Main loop} corresponds to \system's serving phase. \ding{195} \emph{All} includes all code (except NSS) from
the time a process is spawned (used by \sf).

During the experiment, we discovered a variety of suspicious
differences with \tsp. Upon thorough investigation, %
we found a series of issues in \tsp that are rooted in coding bugs,
incomplete algorithms, and its inherent inability to handle
inline assembly. We confirmed all the problems by consulting with the authors
of \tsp~\cite{temporal} and list all of them below.  Because of them, \tsp can
miss required system calls or include unused ones. To obtain a more representative
comparison, we fixed all problems aside \#\ref{prob11}, which requires modifications to \tsp's
algorithm. We use \tspfixed to refer to the version of \tsp with all our fixes incorporated.
\paragraph{\tsp Issues Leading to Missing Required System Calls} (refer to Appendix~\ref{sec:a1} for more details): \
\begin{enumerate}[widest=11] %
	\item \label{prob2} {Cannot handle certain inline assembly constructs.}

	\item \label{prob1} {Does not handle \code{syscall} function.}

	\item \label{prob3} {Does not handle certain function aliases in glibc.}

	\item \label{prob4} {Ignores AT functions in initialization and dynamic-linker
		code.}

	\item \label{prob5} {Does not handle function pointers passed to library
		functions.}

	\item \label{prob6} {Does not handle glibc NSS libraries ~\cite{nss}.}

	\item \label{prob7} {Library not considered due to missing CFG.}

\end{enumerate}

\paragraph{\tsp Issues Leading to Additional Allowed System Calls} (Refer to Appendix ~\ref{sec:a2} for more details):%
\begin{enumerate}[widest=11]\setcounter{enumi}{7} \item \label{prob8}
	{Source-code analysis (Egypt) cannot differentiate between multiple libraries
		in the same directory and merges all of them in the resulting FCG.}

	\item \label{prob9} {Implementation erroneously considers certain null nodes in glibc callgraph as actual functions and uses them during analysis.}

	\item \label{prob10} {TSP includes all exported functions as AT functions in
		cases where callgraph of libraries are not present.}

	\item \label{prob11} {Algorithm fails to prune certain inaccessible AT
		functions.} %

\end{enumerate}

Table~\ref{tab:syscall} summarizes the comparison results.
In summary, \system performs closely to \tsp. On average, we allow 8.33\% more
system calls compared to \tsp. We manually inspected the differences and found that they
are mainly attributed to the availability of source code to \tsp. With
source code, \tsp can perform point-to analysis to resolve indirect calls
more accurately (more details in section ~\ref{sec:a3}). In contrast, \system only assumes binary code and our VFA powered FCG refining still results in an inflated function-call graph. In the last column of ~\autoref{tab:syscall}, we list
the number of system calls that are reachable from AT functions through direct edges.
It indicates that our approach represents the best efforts unless
better methods that can track values through global and heap memory are available to resolve indirect calls in binary code.

\begin{table}[ht!]
    \small
    \centering
	\caption{Mitigated kernel vulnerabilities. \SF and \textbf{SP} stand for \sf and \system, respectively.
		\label{tab:kernelcve}}
		\vspace{-1em}
	\begin{tabular}{l|l|c|c|c}
		\toprule
		\multirow{1}{*}{\textbf{CVE}} & \textbf{System Calls} &  \multirow{1}{*}{\textbf{\SF}} &  \multirow{1}{*}{\textbf{SP}} & \textbf{\tsp}/\textbf{\tspfixed}\\
		\midrule
		\multirow{2}{*}{2018-18281} & execve(at), & \multirow{2}{*}{0} & \multirow{2}{*}{0} & \multirow{2}{*}{4/4} \\
         & mremap &  &  & \\
		\hline
		2016-3672 & \multirow{10}{*}{execve(at)} & \multirow{10}{*}{0} & \multirow{10}{*}{2} & \multirow{10}{*}{4/4} \\
		2015-3339 &  &  &  &  \\
       2015-1593 & & & & \\
        2014-9585 & & & & \\
        2013-0914 & & & & \\
        2012-4530 & & & & \\
        2010-4346 & & & & \\
        2010-3858 & & & & \\
        2008-3527 & & & & \\
        2018-14634 & & & & \\
		\hline
		2012-3375 & epoll\_ctl & 0 & 0 & 1/1 \\
		\hline
        \multirow{3}{*}{2011-1082} & epoll\_ctl,	& \multirow{3}{*}{0}	& \multirow{3}{*}{0}	& \multirow{3}{*}{1/1}  \\
         & epoll\_pwait, &	& 	&   \\
         & epoll\_wait	& 	& 	&   \\
        \hline
        \multirow{1}{*}{2010-4243}	& uselib, execve(at) &\multirow{1}{*}{0}	& \multirow{1}{*}{2}	& \multirow{1}{*}{4/4} \\
        \hline
        \multirow{2}{*}{2019-11815}	& clone	& \multirow{2}{*}{0} &	\multirow{2}{*}{2}	& \multirow{2}{*}{1/2} \\
        & unshare & & &  \\
        \hline
        2013-1959	& write &	0	& 0	& 0/0 \\
        \hline
        2015-8543	& socket &	0	& 0	& 0/0 \\
        \hline
        \multirow{2}{*}{2017-17712}	& sendto	& \multirow{2}{*}{0}	& \multirow{2}{*}{0} &	\multirow{2}{*}{0/0} \\
        & sendmsg & & &  \\
        \hline
        \multirow{2}{*}{2013-1979}	& recvfrom & \multirow{2}{*}{0} &	\multirow{2}{*}{0}	& \multirow{2}{*}{0/0} \\
        & recvmsg & & & \\
		\hline
        2016-4998 & \multirow{3}{*}{setsockopt} & \multirow{3}{*}{0} & \multirow{3}{*}{1} & \multirow{3}{*}{0/0} \\
		2016-4997 &  &  &  &  \\
		2016-3134 &  &  &  &  \\
		\hline
        \multirow{2}{*}{2017-18509}	& setsockopt, & \multirow{2}{*}{0} &	\multirow{2}{*}{1}	& \multirow{2}{*}{0/0} \\
        	& getsockopt &  &	&  \\
		\hline
    	2017-14954 & waitid & 6 & 6 & 6/6 \\
        \hline
		2014-5207 & mount & 6 & 6 & 6/6 \\
		\hline
        2018-12233 & setxattr & 6 & 6 & 6/6 \\
        \hline
		2016-0728 & \multirow{2}{*}{keyctl} & \multirow{2}{*}{6} & \multirow{2}{*}{6} & \multirow{2}{*}{6/6} \\
    	2014-9529 &  &  & &  \\
        \hline
        2019-13272 & \multirow{2}{*}{ptrace} & \multirow{2}{*}{6} & \multirow{2}{*}{6} & \multirow{2}{*}{6/6} \\
		2018-1000199  &  &  & &  \\
        \hline
		\multirow{1}{*}{2014-4699} & fork, clone, ptrace & \multirow{1}{*}{0} & \multirow{1}{*}{2} & \multirow{1}{*}{1/2} \\
        \hline
		2014-7970 & pivot\_root & 6 & 6 & 6/6 \\
		\hline
		2019-10125 & io\_submit & 6 & 6 & 6/6 \\
		\hline
		2017-6001 & perf\_event\_open & 6 & 6 & 6/6 \\
		\hline
		2016-2383 & bpf & 6 & 6 & 6/6 \\
		\hline
		2018-11508 & adjtimex & 6 & 6 & 6/6 \\
		\bottomrule
	\end{tabular}
\end{table}

\subsection{Security Benefits}

\subsubsection{Kernel Attack Surface Reduction:}

This experiment evaluates the effectiveness of \system in reducing kernel
attack surface using the 36 kernel vulnerabilities also used by prior
work~\cite{sysfilter,temporal}. We count the number of applications where \sf,
\tsp, and \system filter the system calls required to trigger each
vulnerability. The evaluation results are summarized
in~\autoref{tab:kernelcve}. \system performs either better or the same as \sf,
regardless of which vulnerability we consider. Compared to \tsp, \system
performs better or the same for 24 vulnerabilities but worse in 12
\code{execve}-based ones. Fundamentally, \system allows \code{execve} in those
cases because it is reachable from AT functions stored in global data. Our VFA
cannot track their propagation and, thus, cannot rule them out from indirect
call targets.

\subsubsection{Exploit Mitigation and Hindrance:}

\begin{table}[ht!]
	\caption{List of equivalent system calls.\label{tab:eqsys}}
    \scalebox{0.9}{
	\begin{tabular}{ll}
		\toprule
		\code{execve} & \code{execveat} \\
		\code{accept} & \code{accept4} \\
		\code{dup} & \code{dup2}, \code{dup3} \\
		\code{eventfd} & \code{eventfd2} \\
		\code{chmod} & \code{fchmodat} \\
		\code{recv} & \code{recvfrom}, \code{read} \\
		\code{send} & \code{sendto}, \code{write} \\
		\code{open} &  \code{openat} \\
		\code{select} &
		\parbox[t]{185pt}{
			\raggedright
			\code{pselect6}, \code{epoll{\_}wait},
			\mbox{\code{epoll{\_}wait{\_}old}},
			\code{poll}, \code{ppoll},
			\code{epoll{\_}pwait}
		}\\
		\bottomrule
	\end{tabular}
    }%
\end{table}

We also evaluate the effectiveness of \system in thwarting exploit shellcode.
Specifically, we reuse the 535 shellcodes involved in the evaluation of \tsp
and consider a shellcode to be stopped if at least one needed system call is
filtered. Considering that certain Linux system calls provide interchangeable
functionalities and adversaries could easily adapt the shellcode to use
alternative system calls, we deem a payload to be stopped only when all
equivalent system calls are filtered. The list is shown in Table
~\ref{tab:eqsys}. To identify equivalent system calls, we follow the same
rules as proposed in the \tsp paper~\cite{temporal}.

\begin{table*}
    \centering
    \caption{ Percentage of shellcodes (total 535) stopped by \system
    (\textsc{SP}), \sf (\SF), and \tsp (original and after
    fixes). Subscripts \textit{With} and \textit{Without} indicate whether we
    consider equivalent system calls or not, respectively.}
    \vspace{-0.5em}
    \label{tab:allpayloads}
    \begin{tabular}{c|cccc|cccc}
    \toprule
    \textbf{Application} &  \textbf{\textsc{SF$_{With}$}} & \textbf{\textsc{SP$_{With}$}} & \textbf{\textsc{TSP}$_{With}$} &  \textbf{\tspfixed$_{With}$} & \textbf{\textsc{SF$_{Without}$}} & \textbf{\textsc{SP$_{Without}$}} &       \textbf{\textsc{TSP}$_{Without}$} &  \textbf{\tspfixed$_{Without}$}\\
    \midrule
    Bind &  20.37 & 67.85 & 67.47 &  71.96 & 33.45 & 71.40 & 76.82 &	73.45 \\
    Httpd & 36.82 & 41.68 &  78.13 & 78.13 & 46.16 & 52.89 &  83.92 & 83.92 \\
    Lighttpd &  32.14 & 58.69 &  60.37 & 60.37 &  34.01 & 60.37 &  62.05 &  62.05 \\
    Memcached &  16.63 & 77.1 & 72.89 & 73.08 & 38.50 & 78.50 & 74.39 &  74.57 \\
    Nginx &  12.89 & 30.28 &  45.42 & 47.85 & 34.95 & 53.83 &  68.22 & 	68.41 \\
    Redis &   33.08 & 49.53 & 49.15 & 49.53 & 38.31 & 65.23 &  65.98 & 	65.23\\
    \bottomrule
    \end{tabular}
\end{table*}

We summarize the results in~\autoref{tab:allpayloads}.
Unsurprisingly, \system outperforms \sf by a large margin but
stops fewer shellcodes than \tsp for most applications (again because of
the \code{execve} system call). In the case of \memcached, \system
outperforms \tsp. The reason is that \system can filter out
\code{getsockopt} and \code{setsockopt}, while \tsp cannot due to
issue~\ref{prob11} discussed in \S\ref{sec:filtered}.

\def\critsysangle{55}
\begin{table}
	\centering
	\caption{Security-sensitive system calls filtered by \system, \tspfixed, and \sf. \label{tab:critcalls}}
	\vspace{-1.5em}
	\begin{tabular}{c c c c c c c}
		\textbf{Syscall} & \rotatebox{\critsysangle}{\textbf{\bind}} & \rotatebox{\critsysangle}{\textbf{\httpd}} & \rotatebox{\critsysangle}{\textbf{\lighttpd}} & \rotatebox{\critsysangle}{\textbf{\memcached}} & \rotatebox{\critsysangle}{\textbf{\nginx}} & \rotatebox{\critsysangle}{\textbf{\redis}} \\
        \toprule
        accept & \xmark & \cmark & \xmark & \xmark & \xmark & \xmark \\
        accept4 & \xmark & \xmark & \xmark & \xmark & \xmark & \cmark \\
        bind & \xmark & \faCircleO & \faCircleO & \xmark & \xmark & \xmark \\
        chmod & \faCircleO & \cmark & \faAdjust & \cmark & \xmark & \faAdjust \\
        clone & \faAdjust & \xmark & \xmark & \xmark & \faAdjust & \xmark \\
        connect & \xmark & \xmark & \xmark & \xmark & \xmark & \xmark \\
        execve & \faAdjust & \faCircleO & \xmark & \faAdjust & \faCircleO & \xmark \\
        execveat & \cmark & \cmark & \cmark & \cmark & \cmark & \cmark \\
        fork & \cmark & \cmark & \cmark & \cmark & \cmark & \cmark \\
        listen & \xmark & \faCircleO & \faAdjust & \xmark & \faCircle & \faAdjust \\
        mprotect & \xmark & \xmark & \xmark & \xmark & \xmark & \xmark \\
        ptrace & \cmark & \cmark & \cmark & \cmark & \cmark & \cmark \\
        recvfrom & \faCircle & \faAdjust & \xmark & \xmark & \xmark & \faAdjust \\
        setgid & \faCircleO & \faCircleO & \faAdjust & \faAdjust & \xmark & \faAdjust \\
        setreuid & \faAdjust & \faAdjust & \faAdjust & \faAdjust & \faAdjust & \faAdjust \\
        setuid & \faCircleO & \faCircleO & \faAdjust & \faAdjust & \xmark & \faAdjust \\
        socket & \xmark & \xmark & \xmark & \xmark & \xmark & \xmark \\
		\bottomrule
	\end{tabular}
	\\ \raggedright
	\xmark: Filtered by none; \cmark: Filtered by all; \faAdjust: Filtered by all but \sf; 
	\faCircleO: Filtered only by \tspfixed; \faCircle: Filtered only by \system. \\
\end{table}

\paragraph{Security-Sensitive System Calls}

We further narrow down our attention to security-sensitive systems calls
(or in \tsp's definition, system calls frequently used in payloads because
of their usefulness to attackers).~\autoref{tab:critcalls} presents our findings.
Out of 102 cases (17 system calls in 6 applications), \tsp outperforms \system
in 10 cases (9.8\%). This is due to the over-approximation of indirect
call targets by \system and the gains delivered to \tsp by its points-to
analysis. In two cases (~2\%), \system
outperforms \tsp. Specifically, \code{listen} in \nginx and \code{recvfrom} in
\bind are \emph{not} filtered by \tsp due to issue~\ref{prob11}.
In all other cases, \system and \tsp are similar. Compared to \sf,
\system performs better in 24 cases (23.5\%).

\begin{table*}[ht!]
\caption{Results of dynamic library analysis. For each application, we list the
configuration tested and the number of \code{dlopen()}/\code{dlsym()} call
sites that we statically resolved \emph{fully}, \emph{partially}, or \emph{not
at all}. The numbers in parentheses indicate the call sites that were also
observed dynamically. We also list the number of additional syscalls added by
the analysis for each configuration. The footnotes annotate the reasons why
full resolution is not achieved.
}
\label{tab:dlopendlsym}
\centering
\begin{tabular}{l | l | p{25pt} p{25pt} p{25pt} | p{25pt} p{25pt} p{25pt} | c}
\toprule
\textbf{Application} & \textbf{Configuration} & \multicolumn{3}{c|}{\textbf{dlopen()}} & \multicolumn{3}{c|}{\textbf{dlsym()}} & \textbf{+Syscalls} \\
    & & Full & Partial & Unres. & Full & Partial & Unres. & \\
\midrule

\multirow{2}{*}{\bind} & Default			& 4 (3)\astiv & 0 (0) & 1 (0)\astiii & 3 (3) & 0 (0) & 2 (0)\astiii & 0 \\
                        & +\code{DLZ module}	& 4 (3)\astiv & 0 (0) & 1 (0)\astiii & 3 (3) & 0 (0) & 2 (0)\astiii & 0 \\
\hline

\multirow{2}{*}{\httpd} & Default			& 0 (0) & 0 (0) & 1 (1)\astii & 0 (0) & 0 (0) & 1 (1)\astii & 3 \\
                        & +\code{mod\_ssl}	& 0 (0) & 0 (0) & 1 (1)\astii & 0 (0) & 0 (0) & 1 (1)\astii & 14 \\
\hline

\multirow{2}{*}{\lighttpd}				& Default & 0 (0) & 0 (0) & 1 (1)\astii & 0 (0) & 0 (0) & 1 (1)\astii & 1 \\
                        & +\code{mod\_(cgi)} & 0 (0) & 0 (0) & 1 (1)\astii & 0 (0) & 0 (0) & 1 (1)\astii & 1 \\
\hline

\memcached				& Default (w/\code{SASL}) & 0 (0) & 1 (1)\astiii\astiv & 0 (0) & 0 (0) & 1 (1)\astii\astiii & 0 (0) & 13 \\

\hline
\multirow{2}{*}{\nginx} 	& Default			& 2 (1)\astiv & 0 (0) & 1 (0) \astiii & 3 (3) & 0 (0) & 2 (0) \astiii & 0 \\
                        & +\code{ngx\_http\_image\_filter\_module}	& 2 (1)\astiv & 0 (0) & 1 (0) \astiii & 3 (3) & 0 (0) & 2 (0) \astiii & 20 \\

\hline
\multirow{2}{*}{\redis} 	& Default			& 1 (1)\astiv & 0 (0) & 0 (0) & 3 (1) & 0 (0) & 0 (0) & 0 \\
                        & +\code{redis\_cell}	& 1 (1)\astiv & 0 (0) & 0 (0) & 3 (1) & 0 (0) & 0 (0) & 0 \\
\bottomrule

\multicolumn{9}{l}{\astii Read from configuration file. \astiii Limitations
of VFA. \astiv Through heuristic based on \code{dlsym()} resolved symbols.}
\end{tabular}
\end{table*}

\subsection{Dynamically Loaded Libraries}
\label{sec:dyneval}

We propose dynamic library profiling in~\S\ref{sec:dynamic} to resolve the
libraries loaded at runtime. In this evaluation, we measure the accuracy and the necessity of this approach.
We run the servers with both the default configuration and customized configurations
where different modules are enabled. For \redis and \lighttpd, we use the test cases
shipped with the respective packages. For \nginx and \httpd, we use \code{nginx-tests}~\cite{nginxtest}
and \code{Apache-Test-1.4.3}~\cite{apachtest} as test cases, respectively.

By analyzing the outcomes of the evaluation above, we unveil that the arguments to \code{dlopen()} and \code{dlsym()} are either hardcoded, read from a configuration file, or constructed dynamically by concatenating strings. In ~\autoref{tab:dlopendlsym}, we show how well we can resolve those arguments. Given \redis and \memcached, our static analysis---combing VFA and heuristics---can resolve all \code{dlopen()} and \code{dlsym()}. For \bind and \nginx, we can resolve all but one \code{dlopen()} and two \code{dlsym()}. The unresolved cases are due to limitations of VFA when handling dynamically generated arguments in the \code{libcrypto} library. These callsites are not observed during runtime too. In the cases of \httpd and \lighttpd, our analysis cannot resolve any case because the arguments are loaded from configuration files. \system presents a better accuracy in resolving dlopen and dlsym arguments when compared with \sf. Out of 12 \code{dlopen()} cases, \system  can resolve 8 while \sf resolves none. Out of 16 \code{dlsym()} cases, \system can handle 9, but \sf only tackles 6.

We further count the system calls used by the dynamically loaded libraries and explore their impacts. In four applications (\httpd, \lighttpd, \memcached, and \nginx), the dynamic libraries require additional system calls. In those cases, it is essential to resolve the dynamic libraries. Otherwise, the system-call filter can break the normal functionality. As expected, taking the dynamically loaded libraries into account can reduce the security benefits. By further allowing the system calls listed in~\autoref{tab:dlopendlsym}, we observe a 2.77\% drop in kernel vulnerability mitigation on all configurations of \httpd and \lighttpd and a 5.47\% drop when considering \memcached with the default configuration.

{\color{\revcolor}

\subsection{\texttt{execve} System Call}
During the evaluation, \system statically identifies that
\texttt{execve} is used to launch ``/bin/sh'' by \httpd and \lighttpd. The binary or shell script
further executed by
``/bin/sh'' cannot be resolved statically. \system also finds that
\texttt{execve} is used in \nginx and \redis while the arguments
cannot be determined statically. Further running dynamic analysis under the
default configurations, however, \system observes no use of \texttt{execve}
by those four programs.

After a closer look via manual analysis, we find that our static analysis is not wrong.
The four programs only invoke \texttt{execve} under special configurations. Specifically:

\begin{itemize}
\item \lighttpd: when configured to load CGI modules via \code{mod\_cgi}, it invokes
\texttt{execve} to run CGI scripts.

\item \httpd: it can also be configured to launch CGI scripts via \texttt{execve}.

\item \nginx: when requested to upgrade the server binary upon special signals,
it uses \texttt{execve} to launch a new version of the server. This occurs outside the
main loop\footnote{\label{execvefootnote}\color{\revcolor} \system cannot filter \texttt{execve} in this case because
the function calling \texttt{execve} is determined as an AT function. Due to approximation,
\system considers that the AT function can be called by indirect calls inside the main loop.} and does not impact our filter.

\item \redis: when configured to run in the debug mode, it uses
\texttt{execve} to restart the server upon request. Also, when configured to run in the sentinel mode, it uses \texttt{execve} to run pending scripts. These occur outside of the main loop\cref{execvefootnote} and does not happen in the default release mode. Thus, it has no impact on our filter.
\end{itemize}

To sum up, servers can use \texttt{execve} under certain configurations.
To determine the arguments of \texttt{execve} in those cases, \system will
need the users' help with setting up the desired configurations for its dynamic analysis.

\subsection{Robustness and Efficiency}

\subsubsection{Compilers and Optimizations:} The compiler and optimizations
used to build an application affect the resulting binary. We evaluate the effects
those on \system, by building our benchmarks using both GCC-7.5.0 and CLANG-6.0.0
under varying optimization settings. We apply
\system on the resulting binaries and count the number
of allowed syscalls in the main loop. The evaluation
results are presented in~\autoref{tab:olevels}.

Overall, GCC and CLANG lead to nearly identical results.
Only in the case of \redis, \system identifies one more syscall
given GCC-compiled binaries. Throughout further analysis,
we find that this is because GCC (since version 4.0) enforces
\code{-D\_FORTIFY\_SOURCE=1} at optimization level O1 and above~\cite{fortify}.
This will transform \code{longjmp} to a checked version \code{\_\_longjmp\_chk},
and the latter additionally needs the \code{sigaltstack} syscall. In contrast,
CLANG does not set up \code{-D\_FORTIFY\_SOURCE} by default, avoiding the use of
\code{sigaltstack}. We want to note that \redis compiled by GCC under O0
does not include \code{sigaltstack}. However, it presents one extra dummy
AT function, which leads to the inclusion of the \code{creat} syscall. This
is why that binary under O0 has the same amount of syscalls.

The optimization levels present no impact except for
the cases of \memcached and \redis. Given \memcached compiled under
optimization level O0 (with both GCC and CLANG), \system
detects 13 more syscalls that are needed for the main loop.
The reason---based on our inspection---is that the O1-O3 optimization
levels more aggressively eliminate unreachable code, inline functions,
and unroll loops, etc. These lead to a more accurate analysis by
\system (in particular fewer AT functions) and thus, reduce the number of
dummy syscalls. The detail about \redis has been discussed above.

\def\benchangle{60}
\begin{table}[ht]
	\centering
	\caption{\color{\revcolor}
    Allowed system calls when building the benchmarks with
	different compilers and optimization levels. The underlined values correspond to the configurations tested in \S\ref{sec:filtered}.
    \label{tab:olevels}}
	\vspace{-1.5em}
	\begin{tabular}{c c l l l l l l}
		& & \rotatebox{\benchangle}{\textbf{\bind}} & \rotatebox{\benchangle}{\textbf{\httpd}} & \rotatebox{\benchangle}{\textbf{\lighttpd}} & \rotatebox{\benchangle}{\textbf{\memcached}} & \rotatebox{\benchangle}{\textbf{\nginx}} & \rotatebox{\benchangle}{\textbf{\redis}} \\
        \toprule
        \multirow{4}{*}{\rotatebox{90}{\textbf{GCC}}}
        & O0 & 103 		& 92 		& 93		& 98		& 104		&  85 \\
        & O1 & 103 		& 92 		& 93		& 85		& \ul{104}		& 85 		\\
        & O2 & \ul{103} & \ul{92}	& \ul{93}			& \ul{85}	& 104	& \ul{85}	\\
        & O3 & 103 		& 92 		& 93 				& 85 		& 104	& 85		\\
        \midrule
        \multirow{4}{*}{\rotatebox{90}{\textbf{CLANG}}}
        & O0 & 103 & 92 & 93 & 98 & 104 & 84 \\
        & O1 & 103 & 92 & 93 & 85 & 104 & 84 \\
        & O2 & 103 & 92 & 93 & 85 & 104 & 84 \\
        & O3 & 103 & 92 & 93 & 85 & 104  &  84\\
		\bottomrule
	\end{tabular}
\end{table}

\subsubsection{Binary-only Software:} To evaluate \system, we use open-source software
so that we can verify the correctness of the results. To further assess its ability to
handle binaries, we apply \system on a proprietary, closed-source web server, the
Abyss Web Server v2.16 from Aprelium~\cite{Abyss}.
Unlike modern binaries, Abyss is not compiled as position-independent
code (PIC)---which is necessary for benefiting from basic defenses like ASLR.
This forces further over-approximation of
AT functions and possibly inflates the number of required syscalls.
This also breaks Egalito's functionality to correctly rewrite Abyss and inject the
code setting up the Seccomp-BPF filter.
To inject the filter, we use another static binary rewriter \epatch~\cite{e9patch:pldi20}.
\epatch minimally alters the Abyss binary to add a trampoline to the
Seccomp-BPF filter program before the main loop entrance. To make it
work, we include the definition of \code{prctl()} to the libc tailored
by \epatch. To run dynamic analysis, we profile Abyss with the
default configurations and 10k HTTPD requests.

The evaluation finds that Abyss spawns six
threads at startup for different services.
\system detects a main loop for each thread,
and we summarize the details in~\autoref{tab:abyss}.
\system reports that the \code{main()} of the
server only requires 86 syscalls for correct operation,
filtering security-sensitive syscalls including
\code{accept4}, \code{chmod}, \code{execveat}, \code{ptrace},
\code{setreuid}, and \code{fork}. Narrowing down
to the serving phase (i.e., the main loop),
\system further filters out 4 more syscalls (\texttt{pipe}, \texttt{dup2},
\texttt{setsid}, and \texttt{execve)} for two threads
and one more syscall (\texttt{setsid}) for the main thread.
To further validate the stability
of the server after the Seccomp-BPF filter is inserted,
we rerun the patched server with the aforementioned 10k HTTP requests.
It shows that the server works without exceptions.

We also inspect the CGI modules of Abyss. When enabled,
similar to \lighttpd and \httpd, Abyss calls \code{execve}
to launch ``/bin/sh'' for executing the CGI scripts. This
explains why \system cannot filter \code{execve} for
Abyss' request-processing thread.

\begin{table}[ht]
        \small 
	\centering
	\caption{\color{\revcolor}
    Main loops detected in the Abyss web server and the number of syscalls
	required by them. Partition 0 is the main process and Partitions 1 to 5 are threads of child process. Thread in partition 3 spawns new threads to handle requests. \code{main()} requires 86 syscalls.}
    \label{tab:abyss}
	\begin{tabular}{ c c c c c}
		\toprule
        \multicolumn{3}{c}{\textbf{Partition Address}} & \textbf{Main Loop} &   \textbf{Filtered}\\
        ID & Function & Main Loop & \textbf{syscalls}  & \textbf{syscalls}\\
		\midrule
        0 & 0x443da0 & 0x443dca & 85  & setsid\\
       1 &  0x42edd0 &  0x42efea & 82 &  pipe,dup2,execve, setsid \\
        2 & 0x493330 &  0x4933e8 & 82 & pipe,dup2,execve, setsid \\
        3 & 0x466740 & 0x466771	& 86  & -- \\
        4 & 0x405960 & 0x405970	& 86 &  -- \\
        5 & 0x459000 & 0x45906a	& 86 & -- \\
		\bottomrule
	\end{tabular}
\end{table}

}%

\subsubsection{Analysis Speed:}

We compare the execution times of the static analysis phase (call-graph
construction and system-call set generation) between \system and \tsp
in~\autoref{tab:exectime}. In all cases, \system runs at least 81\% faster
than \tsp.

\subsubsection{FCG Improvements over Egalito:}

Comparing with Egalito, our FCG refinement using VFA and TypeArmor achieved a
reduction in FCG edges of 8.34\%, 6.68\%, 2.36\%, 3.93\%, 6.74\% and 10.36\%
in \bind, \httpd, \lighttpd, \memcached, \nginx and \redis, respectively.

\begin{table}
\caption{Analysis time (in seconds) of \system and \tsp.}
\vspace{-1em}
\label{tab:exectime}
\begin{tabular}{l | c | c}
\toprule
\textbf{Application}& \textbf{\tsp} & \textbf{\system} \\
\midrule
\bind		 &	562 & 4.06	(-99.27\%) \\
\httpd 		 &	17	& 0.73	(-95.68\%) \\
\lighttpd	 &	3	& 0.56	(-81.10\%) \\
\memcached	 &	3	& 0.45	(-85.00\%) \\
\nginx 		 &	83	& 5.01	(-93.95\%) \\
\redis		 &	23	& 1.23	(-94.63\%) \\
\bottomrule
\end{tabular}
\end{table}

\section{RELATED WORK}
\label{sec:related}

\subsection{System-Call Interposition}

System-call interposition~\cite{janus,ostia,sekar} is an early research direction to restrict system calls.
It interposes at interfaces between the application and the OS kernel to enforce security policies. Janus~\cite{janus}
leverages \texttt{ptrace} to dynamically monitor and restrict system calls that an application can perform.
In contrast, Ostia~\cite{ostia} sandboxes an application and runs a delegate program to make system calls
on behalf of the application according to a user-specified security policy. Besides intercepting system calls
through a pair of kernel database and a user-space daemon, Systrace~\cite{sysinter} further automates the generation of security policies using dynamic analysis.

\subsection{System-Call Filtering}
System-call filtering represents a more recent method to limit system calls. The idea is to identify the set of
system calls required by the application and filter the unneeded ones at runtime through mechanisms like Seccomp-BPF.
Constructing and mounting the filters is trivial. Thus, research in this line primarily focuses on
determining the list of required system calls. Abhaya~\cite{abhayaand}, Chestnut~\cite{chestnut}, Confine~\cite{confine},
and \sf~\cite{sysfilter} all rely on static analysis to over-approximate the set of system calls. Notably,
Abhaya and \sf both consider the dependent libraries, with \sf being binary only. \sf also employs dynamic library profiling using static VFA,  but our VFA with heuristics performs better in resolving \code{dlopen} and \code{dlsym} callsites as mentioned in section ~\ref{sec:dyneval}. In contrast, SIT~\cite{zeng2013}
runs best-effort static analysis to get an initial allow-list of system calls and incorporates runtime monitoring
to compensate for false negatives. BASTION~\cite{syscallintegrity} introduces the concept of ``integrity'' to
system-call filtering. With BASTION, syscall invocations are bound by call type integrity, control-flow integrity and argument integrity. Adopting a similar
principle, Shredder~\cite{mishra2018shredder} derives the expected argument values of system APIs
and specializes the APIs to only allow those values.

\tsp~\cite{temporal} advances system-call filtering to be temporal. Instead of viewing an application as a whole,
\tsp separates the execution of the application into an initialization phase and a serving phase. It tailors different
filter rules for different phases. As described in~\S\ref{sec:tspback}, \tsp presents several major limitations to achieve
practical temporal system-call filtering. Our system \system, overcoming those limitations, offers the first
solution to provide automated, binary-only, and robust temporal system-call filtering.

\subsection{Code Debloating}
Code debloating removes code not required by an application or its libraries during runtime, which can help reduce the attack surface. RAZOR~\cite{razor} utilizes a set of test cases and control-flow-based heuristics to perform code reduction for deployed binaries, preserving only the essential code needed to support user-expected functionalities. Chisel~\cite{chisel} also reduces unneeded code from the application, but using machine learning with a high-level specification of the desired functionality. In contrast, CodeFreeze~\cite{codefreeze}, Nibbler~\cite{nibbler:acsac19}, BlankIt~\cite{blankit}, and configuration-driven software debloating~\cite{c2c} focus on debloating library code, using static analysis and/or dynamic analysis. Specifically, CodeFreeze and Nibbler
remove unused code from dynamic libraries at loading time, while configuration-driven software debloating~\cite{c2c} identifies the required libraries based on the configurations and disable the remaining by not loading them. BlankIt~\cite{blankit} uses a more fine-grained strategy by only loading the library functions that will be used at each call site at runtime. Different from these works, Piece-wise~\cite{piecewise} combines compilation-time analysis and loading-time enforcement to only load code needed by the program, offering support for both applications and libraries. It combines a static and training based approach to resolve arguments to \code{dlopen} and \code{dlsym}, although no evaluation results are provided for this analysis.

\section{CONCLUSION}
We presented \system, an automatic system-call filtering system for
binary-only server programs. \system identifies the serving
phases of applications threads, computes the set of system calls required by
each, and installs an efficient \seccomp filter that disallows other system
calls. We implemented \system on x86-64 Linux and evaluated it on six popular
server applications. The results demonstrate that \system accurately locates
the point where serving phase begins and performs comparably to prior
source-code-based work~\cite{temporal}, but without errors.
Moreover, \system only allows 8.33\% more syscalls overall, while it filters
as many security-critical syscalls in 88.23\% of cases. Unsurprisingly, it
also outperforms prior work~\cite{sysfilter} on binaries that does not
consider execution phases.
In terms of security, \system is successful in blocking exploit payloads and
preventing kernel vulnerabilities with success rates ranging from 53.83\% to
78.5\% and 36.11\% to 77.77\%, respectively, for the ones tested.
Finally, \system surpasses prior work~\cite{sysfilter} in soundly resolving
instances of \code{dlopen()} and \code{dlsym()} by 58.33\% and 18.75\%,
respectively.

\begin{acks}
We thank the anonymous reviewers for their valuable comments and time. This work was supported by
the Office of Naval Research (ONR) awards N00014-17-1-2788, N00014-22-1-2643,
and N00014-17-1-2787, DARPA award D21AP10116-00, and National Science
Foundation (NSF) awards CNS-2213727 and OAC-2319880. Any opinions, findings, and conclusions or
recommendations expressed herein are those of the authors and do not
necessarily reflect the views of the US government, ONR, DARPA, or NSF.
\end{acks}

\bibliographystyle{ACM-Reference-Format}
\balance
\bibliography{references,sysfilter,portokalidis}
\appendix
\newpage
\section{APPENDIX}
\label{sec:appendix}

\subsection{\tsp Issues Leading to Missing Required System Calls}
\label{sec:a1}
\subsubsection{Cannot handle certain inline assembly constructs :}
In glibc, inline assembly code is used to invoke system calls using \code{syscall}
instruction. In most cases, macro \code{INLINE\_SYSCALL} is used to invoke the
system calls which is recognized by TSP. But there are some patterns which are not recognized by \tsp due to which the system calls are missed by \tsp.

\subsubsection{Does not handle \code{syscall()} function :}
System calls can be invoked using the \code{syscall()} function. The first argument
to the \code{syscall()} function is the system call number that identifies the system
call. \tsp detects \code{syscall()} but cannot determine the system call
number and hence misses the corresponding system call.

\subsubsection{Does not handle certain function aliases in glibc :}
Glibc uses internal alias functions. The glibc callgraph produced by \tsp using Egypt tool has callpaths which are incomplete. This causes TSP to miss the following system calls.

 For example, in the glibc callgraph generated by \tsp, \code{luaL\_loadfile()} invokes \code{freopen()}, which invokes \code{*\_\_GI\_\_dup3()}. \code{dup3()} invokes \code{\_\_dup3()} which invokes \code{dup3} system call. But there is no outgoing edge for *\_\_GI\_\_dup3() in \tsp glibc callgraph and hence the path to \code{dup3} system call is missed by TSP in Redis.

\subsubsection{Ignores AT functions in initialization and dynamic-linker code :}
\tsp doesn't consider the functions which are AT in the initialization section of the application or libraries as well as in the dynamic linker. This causes it to miss certain system calls that are reachable from these AT functions.

\subsubsection{Does not handle function pointers passed to library functions :}
In \tsp, the callgraphs of the application and the shared libraries are generated separately. Hence, in cases where there are AT functions which are passed as arguments to shared libraries, the callgraph generation algorithm of TSP doesn't generate any outgoing edges for those AT functions as a result of which the functions that are invoked from those AT functions are missed by TSP. An example is signal handler functions that are passed as argument to \code{sigaction()}. In \nginx, \code{ngx\_signal\_handler} is a signal handler function that is passed as argument to \code{sigaction()}.  \code{ngx\_signal\_handler} has a callpath leading to the system call wait4, which is missed by \tsp.

\subsubsection{Does not handle glibc NSS libraries :}
TSP doesn't resolve NSS functions and hence misses out on the system calls reachable from these functions. An example is \code{sendmmsg} in \lighttpd.

\subsubsection{Library not considered due to missing CFG :}
In certain cases when CFG of a library is not already generated, \tsp tries to find the exported symbols of the library. But if the library is not present in the system at all, then it misses out on all the system calls which are reachable from functions in that library.

\begin{table*}[ht!]
\caption{All main loops detected by \system, including the ones for auxiliary threads/processes.}
\label{tab:allpartitions}
\begin{tabular}{c|c|c}
\toprule
\multirow{2}{*} \textbf{Application} & \textbf{Server in} & \textbf{Partitions detected by} \\
& \textbf{Default config} & \textbf{\system (No: of Syscalls)} \\
\midrule
\multirow{5}{*} \textbf{Bind} & Single-process &\loopfunc{main + 0x13701} (103) \\
& Multi-threaded & \loopfunc{netthread + 0x105} (112) \\
& & \loopfunc{nm\_thread + 0x99} (112) \\
& & \loopfunc{run + 0x358\astii} (112) \\
& & \loopfunc{run + 0x119\asti} (112) \\
\hline
\multirow{6}{*} \textbf{Httpd} & \multirow{6}{*} \textbf{Multi-process} & \loopfunc{ap\_build\_config + 0x402} (92) \\
& Multi-threaded & \loopfunc{ap\_run\_mpm + 0x103} (92) \\
& & \loopfunc{child\_main + 0x5170} (92) \\
& & \loopfunc{listener\_thread + 0x274} (92)\\
& & \loopfunc{start\_threads + 0x360} (92) \\
& & \loopfunc{worker\_thread + 0x388} (92) \\
\hline
\multirow{2}{*} \textbf{Lighttpd} & Single-process &  \loopfunc{main + 0x306} (93) \\
& Single-threaded & \\
\hline
\multirow{7}{*} \textbf{Memcached} & Single-process &\loopfunc{main + 0x26742} (85) \\
& Multi-threaded & \loopfunc{logger\_thread + 0x100} (82) \\
& & \loopfunc{event\_base\_loop + 0x391} (85)\\
& & \loopfunc{assoc\_maintenance\_thread + 0x50} (82) \\
& & \loopfunc{item\_crawler\_thread + 0x276} (82) \\
& & \loopfunc{lru\_maintainer\_thread + 0x579} (82) \\
& & \loopfunc{slab\_rebalance\_thread + 0x100} (82) \\
\hline
\multirow{2}{*} \textbf{Nginx} & Multiple-process &  \loopfunc{ngx\_master\_process\_cycle + 0x5456} (106) \\
& Single-threaded & \loopfunc{ngx\_worker\_process\_cycle + 0x391} (104) \\
\hline
\multirow{2}{*} \textbf{Redis} & Single-process &  \loopfunc{aeMain + 0x22} (85) \\
& Multi-threaded & \loopfunc{bioProcessBackgroundJobs + 0x581} (85) \\

\bottomrule
\multicolumn{3}{l}{\astii: \code{isc/timer.c}, \asti: \code{isc/task.c} }
\end{tabular}
\end{table*}

\subsection{\tsp Issues Leading to Additional Allowed System Calls}
\label{sec:a2}

\subsubsection{Source-code analysis (Egypt) cannot differentiate between
multiple libraries in the same directory and merges all of
them in the resulting glibc FCG :}

The glibc callgraph produced by Egypt in \tsp includes functions from unused modules. This results in including callpaths which are not necessarily accessible. For example,
glibc contains malloc/memusage.c, within which function interposition is used to invoke \code{ malloc(), calloc(), free()} etc for memory profiling. This is
compiled into \code{libmemusage.so}, and the functions within this shared library can be invoked only if this library is preloaded by setting the environmental variable \code{LD\_PRELOAD}. \tsp callgraph contains this callpath, \\
\code{malloc()->me()->creat64()->creat()-> syscall(85)} \\
This causes the system call \code{creat} to be included by \tsp whenever \code{malloc()} is used by an application or shared library, while this callpath will only be invoked if libmemusage.so is enabled, which is not used by any of the servers.

\subsubsection{Implementation erroneously considers certain null nodes
as actual functions and uses them during analysis :}

In the glibc callgraph used by \tsp there are outgoing NULL edges and incoming NULL edges. This cause their traversal algorithm to traverse parts of glibc callgraph which are not actually reachable.

System calls \code{arch\_prctl} and \code{set\_tid\_address} are included by \tsp in all servers due to the following paths from the glibc callgraph.

\code{\_pthread\_cleanup\_push} -> \code{NULL} \\
\code{NULL} ->  \code{\_\_pthread\_initialize\_minimal\_internal} \\
\code{\_\_pthread\_initialize\_minimal\_internal} -> \code{\_\_libc\_setup\_tls} \\
\code{\_\_libc\_setup\_tls} -> \code{arch\_prctl (syscall)} \\

\code{\_pthread\_cleanup\_push} -> \code{NULL} \\
\code{NULL} ->  \code{\_\_pthread\_initialize\_minimal\_internal} \\
\code{\_\_pthread\_initialize\_minimal\_internal} -> \code{set\_tid\_address (syscall)}

\subsubsection{TSP including all exported functions as AT functions in
cases where callgraph of libraries are not present :}

If the callgraph of libraries are not present, \tsp includes all exported functions within the library as AT functions.

\subsubsection{Inefficient Pruning Algorithm :}

TSP employs a pruning algorithm to prune out function pointers which are address taken in functions which are inaccessible from main. But the list of AT functions that are used as input for this algorithm is incomplete and doesn't include all the AT functions used to generate the callgraph, as a result of which the algorithm misses out on pruning certain AT functions which are not reachable from main, and thereby including the system calls from these AT functions.

\subsection{Pruning of Indirect-Call Targets in TSP :}
\label{sec:a3}

\tsp uses SVF implementation of Andersen's algorithm to produce callgraphs of applications. SVF Andersen's considers the number of arguments while resolving indirect call targets. \tsp further refines the callgraph by matching the indirect callsites with functions based on argument types, more specifically struct argument types. Also, it prunes those indirect call edges to those AT functions which are address taken in paths which are inaccessible from main.
These pruned edges causes \tsp to remove some system calls which are accessible using these edges.

\subsection{All Main Loops Detected by \system}

\autoref{tab:allpartitions} lists all the main loops detected by \system.

\subsection{\textcolor{\revcolor}{Extensions to Egalito}}
\label{sec:a5}
\textcolor{\revcolor}{Besides identifying and fixing bugs, we incorporated the following extensions 
into Egalito to support our needs.
\begin{enumerate}[widest=11]
\item Added loop identification (\S\ref{sec:findpartition}).
\item Added VFA analysis for the following :
\begin{enumerate}
\item To resolve indirect-branch targets (\S\ref{sec:refining}) for callgraphs. This completely resolved an additional 2.84\% of indirect branches in the evaluated servers (a 4.24\% reduction in FCG edges).
\item To resolve arguments for \code{dlopen}, \code{dlsym}, \code{execve} (\S\ref{sec:dynamic}).
\item To identify thread start functions (\S\ref{sub:sub:thread}).
\end{enumerate}
\item Improved non-returning function detection (\S\ref{sub:sub:non-return}), discovering an average of 41\% more such functions.
\item Added a new pass to insert the syscall filter before the main loop's start (\S\ref{sub:sub:filter}).
\end{enumerate}}

\end{document}